\pdfoutput=1
\RequirePackage{ifpdf}
\ifpdf 
\documentclass[pdftex]{sigma}
\else
\documentclass{sigma}
\fi

\numberwithin{equation}{section}

\begin{document}

\allowdisplaybreaks

\renewcommand{\thefootnote}{$\star$}

\renewcommand{\PaperNumber}{034}

\FirstPageHeading

\ShortArticleName{Structure Theory for Extended Kepler--Coulomb 3D Classical Superintegrable Systems}

\ArticleName{Structure Theory for Extended Kepler--Coulomb\\ 3D Classical Superintegrable Systems\footnote{This
paper is a contribution to the Special Issue ``Superintegrability, Exact Solvability, and Special Functions''. The full collection is available at \href{http://www.emis.de/journals/SIGMA/SESSF2012.html}{http://www.emis.de/journals/SIGMA/SESSF2012.html}}}

\Author{Ernie G.~KALNINS~$^\dag$ and  Willard  MILLER Jr.~$^\ddag$}

\AuthorNameForHeading{E.G.~Kalnins and  W.~Miller~Jr.}

\Address{$^\dag$~Department of Mathematics,  University
of Waikato, Hamilton, New Zealand}
\EmailD{\href{mailto:math0236@math.waikato.ac.nz}{math0236@math.waikato.ac.nz}}

\Address{$^\ddag$~School of Mathematics,  University of Minnesota,
 Minneapolis, Minnesota,
55455, USA}
\EmailD{\href{mailto:miller@ima.umn.edu}{miller@ima.umn.edu}}
\URLaddressD{\url{http://www.ima.umn.edu/~miller/}}

\ArticleDates{Received March 14, 2012, in f\/inal form June 04, 2012; Published online June 07, 2012}

\Abstract{The classical Kepler--Coulomb system in 3 dimensions is well known to be 2nd order superintegrable, with a symmetry algebra that closes polynomially under
Poisson brackets. This polynomial closure is typical for 2nd order superintegrable systems in 2D and for 2nd order systems in 3D with nondegenerate (4-parameter)
potentials.
However the degenerate 3-parameter potential for the 3D extended Kepler--Coulomb system (also 2nd order superintegrable) is an exception,
as its quadratic symmetry algebra doesn't close polynomially.  The 3D 4-parameter  potential for the extended Kepler--Coulomb system is not even 2nd order
superintegrable. However,  Verrier and Evans  (2008) showed  it was 4th order superintegrable, and   Tanoudis and Daskaloyannis (2011) showed that in the
quantum case, if a second 4th order symmetry is added to the generators, the double commutators in the symmetry algebra  close polynomially.  Here, based on the
Tremblay, Turbiner and Winternitz construction,  we  consider an inf\/inite class of classical extended Kepler--Coulomb 3-~and 4-parameter systems indexed by a pair of
rational numbers $(k_1,k_2)$ and reducing to the usual systems when $k_1=k_2=1$. We show these systems to be superintegrable of arbitrarily high order and work
out explicitly the structure of the symmetry algebras determined by the 5 basis generators we have constructed. We demonstrate that the symmetry algebras close
rationally; only for systems admitting extra discrete symmetries is polynomial closure achieved. Underlying the structure theory is the existence of raising and
lowering constants of the motion, not themselves polynomials in the momenta, that can be employed to construct the polynomial symmetries and their structure relations.}


\Classification{20C35; 22E70; 37J35; 81R12}

\renewcommand{\thefootnote}{\arabic{footnote}}
\setcounter{footnote}{0}

\section{ Introduction}\label{int1}

A  quantum superintegrable system  is an integrable
 $n$-dimensional Hamiltonian system with
 Schr\"odinger operator
\[H=\Delta_n +V({\bf x}),
\]
where $\Delta_n=\frac{1}{\sqrt{g}}\sum_{j,k=1}^n\partial_{x_j}(\sqrt{g}g^{jk})\partial_{x_k}$ is the Laplace--Beltrami operator on a Riemannian mani\-fold,  in local coordinates
$x_j$, $n\ge 2$. The system is required to admit $ 2n-1$ algebraically independent globally def\/ined
partial dif\/ferential  symmetry operators
\[
S_j,\quad j=1,\dots,2n-1,\quad n\ge 2,
\] with $S_1=H$
and $[H,S_j]\equiv HS_j-S_jH=0$,  apparently the maximum number possible.
The system is  of order $\ell$ if the maximum order of the symmetry operators, other than the  Schr\"odinger operator,  is $\ell$.
 Similarly, a  classical superintegrable system with
 Hamiltonian
\[{\cal H}=\sum g^{jk}p_jp_k+V({\bf x})\]
on  phase space with local coordinates $x_j$, $p_j,$ where $ds^2=\sum g^{jk}dx_j  dx_k$ is an integrable system such that  there are $ 2n-1$ functionally independent
functions polynomial in momenta, (easily provable to be the maximum number possible):
\[{\cal S}_j({\bf p},{\bf x}),\quad j=1,\dots,2n-1,\]
with ${\cal S}_1={\cal H}$,  and  globally def\/ined
such that $\{{\cal S}_j,{\cal H}\}=0$, where
\[
\{{\cal F},{\cal G}\}=\sum_{j=1}^n \left(\frac{\partial {\cal F}}{\partial_{x_j}}\frac{\partial {\cal G}}{\partial_{p_j}}-\frac{\partial {\cal F}}{\partial_{p_j}}\frac{\partial {\cal G}}{\partial_{x_j}}\right)
\]
is the Poisson bracket.
  The system is  of order $\ell$ if the maximum order of the generating constants of the motion is $\ell$. As has been pointed out many times \cite{IMA, SCQS} such systems are of enormous historic and present day practical importance. In essence, superintegrable systems are those Hamiltonian systems that can be ``solved" exactly, analytically and algebraically, without requiring numerical approximation. Superintegrable systems are used as the basis for exact and perturbation methods that underlie planetary motion determination, orbital maneuvering, the periodic table of the elements, the boson calculus
and much of special function theory,~\cite{KMP, KMP2010}.

The key  property that makes a system ``superintegrable''  is that, in contrast to merely integrable systems,  the symmetry algebra generated by the basis
symmetries is nonabelian. This nonabelian structure can be analyzed and used to deduce properties of the system. Thus in the quantum case the irreducible
 representations of the symmetry algebra determine the multiplicities of the degenerate energy eigenspaces and permit  algebraic computation of the eigenvalues.

The classical Kepler--Coulomb system in 3 dimensions is well known to be 2nd order superintegrable, with a symmetry algebra that closes polynomially in
 an $so(4)$-like structure, e.g.~\cite{BH}. This polynomial closure (though not usually a Lie algebra) is typical for 2nd order superintegrable systems in 2D
\cite{KKM2,KKMP2009} and
for 2nd order systems in 3D with nondegenerate (4-parameter) potentials. However the degenerate 3-parameter potential for the 3D extended Kepler--Coulomb system
(also 2nd order superintegrable) is an exception,  as its quadratic symmetry algebra doesn't close polynomially \cite{KKM2007}. We write it in the form
\begin{gather}\label{3Keplerham}
{\cal H}=p_x^2+p_y^2+p_z^2+\frac{\alpha}{r}+\frac{\beta}{x^2}+\frac{\gamma}{y^2},
\end{gather}
where $x$, $y$, $z$ are the usual Cartesian coordinates with conjugate momenta $p_x$, $p_y$, $p_z$ in phase space and $r=\sqrt{x^2+y^2+z^2}$.

 The 3D 4-parameter  extended Kepler--Coulomb system is not even 2nd order superintegrable.  We write it in the form
\begin{gather}\label{4Keplerham}
{\cal H}=p_x^2+p_y^2+p_z^2+\frac{\alpha}{r}+\frac{\beta}{x^2}+\frac{\gamma}{y^2}+\frac{\delta}{z^2}.
\end{gather}
However,  Verrier and Evans \cite{Evans2008a} showed  it was 4th order superintegrable,
and   Tanoudis and Daskaloyannis~\cite{DASK2011} showed in the quantum case that,
 if a second 4th order symmetry is added to the generators, the symmetry algebra  closes polynomially in the sense that all second commutators of the generators can be
expresses as symmetrized polynomials in the generators. (Note that the 3-parameter potential is not just a restriction of the 4-parameter case, because it admits symmetries that are not
inherited from the 4-parameter symmetries.)

 Here we introduce an analog of the TTW construction~\cite{TTW1,TTW2}  and  consider an inf\/inite class of classical extended Kepler--Coulomb 3- and 4-parameter
systems indexed by a pair of rational numbers $(k_1,k_2)$ and reducing to the usual systems when $k_1=k_2=1$. We construct explicitly a set of generators,
show these systems to be superintegrable of arbitrarily high order and determine the structure of the generated  symmetry algebras. We demonstrate that the
symmetry algebras close rationally; only for systems admitting extra discrete symmetries is polynomial closure achieved. Much of the paper is quite technical but,
as this is the f\/irst work devoted to
uncovering the structure of the symmetry algebras of 3D superintegrable systems of arbitrary order, we think it is important to expose the details of computations and
concepts  that later may prove to be routine.

For the 4-parameter system in the case $k_1=k_2=1$, where discrete symmetry is present and a 6th generator is needed to obtain polynomial closure, we work out the 12th order functional relationship between the 6 generators.

In Section~\ref{actionangle} we review the action angle construction that we employ to show that our systems are superintegrable and to enable the
determination of the structure of the symmetry algebra. In Section~\ref{3d3k} we use the fact that the 3-parameter Kepler--Coulomb system~(\ref{3Keplerham})
 separates in spherical coordinates $r$, $\theta_1$, $\theta_2$ and, by replacing the angles by $k_1\theta_1$, $k_2\theta_2$ for $k_1$, $k_2$ rational, def\/ine an
 inf\/inite family of extended  Kepler--Coulomb systems, no longer restricted to f\/lat space. We demonstrate that each of these systems is superintegrable,
 but of arbitrarily high order. We use our method of raising and lowering symmetries to determine the structure of the symmetry algebras generated by these systems.
 The general construction does not yield generators of minimum order and we show in Section~\ref{mog3} how a limit argument exposes the minimum order generators.
 Underlying the structure theory is the existence of raising and lowering constants of the motion, not themselves polynomials in the momenta, that can be
 employed to construct the polynomial symmetries and their structure relations. We show that the symmetry algebra closes rationally, not polynomially.

  In Section~\ref{3d4k} we apply our method to the 4-parameter Kepler--Coulomb system (\ref{4Keplerham}) and use its separation in spherical coordinates $r$, $\theta_1$, $\theta_2$. By replacing the angles by $k_1\theta_1$, $k_2\theta_2$ for $k_1$, $k_2$ rational, we def\/ine an inf\/inite family of extended  Kepler--Coulomb systems, again not restricted to f\/lat space. We demonstrate that each of these systems is superintegrable,  but of arbitrarily high order. We use raising and lowering symmetries to determine the structure of the symmetry algebras generated by these systems. Again the general construction does not yield generators of minimum order and we show in Section~\ref{mog4} how a limit argument exposes the minimum order generators.  In general, the symmetry algebra closes rationally, but not polynomially. However in two cases $k_1=k_2=1$ and $k_1=k_2=1/2$ the system admits additional symmetry: the 6-element permutation group~$S_3$. The general construction shows that these systems admit 5 generators of orders  2, 2, 2, 2, 4, but in Section~\ref{k1k2} we show that the permutation symmetry implies the existence of a 6th symmetry of order 4 such that the 6 symmetries are linearly independent. Then we demonstrate that the algebra generated by these 6 symmetries closes polynomially, in analogy with the computation for the quantum case in~\cite{DASK2011}. We go further and work out the 12th order functional relation between the 6 generators.

In Section~\ref{Final Comments} we present our conclusions and prospects for additional research.

\section{Review of the  action-angle construction}\label{actionangle}

In \cite{KKM10, KKM10a,KKM10b,KMPog10, Tsiganov2008} it was described how to determine a complete set of $2n-1$ functionally independent constants of the motion for a
classical Hamiltonian on an $n$-dimensional Riemannian or pseudo-Riemannian manifold whose Hamilton--Jacobi equation separates in an orthogonal
subgroup coordinate system.
In the special case $n=3$ the def\/ining equations for the Hamiltonian~$\cal H$, expressed in the separable coordinates $q_1$, $q_2$, $q_3$, take the form
\begin{gather*}
{\cal H}={\cal L}_1=p_1^2+V_1(q_1)+f_1(q_1){\cal L}_2,\nonumber\\
  {\cal L}_2=p_2^2+V_2(q_2)+f_2(q_2){\cal L}_3,\qquad {\cal L}_3=p_3^2+V_3(q_3). 
\end{gather*}
The additional constants of the motion can be constructed as
\[
{\cal L}'_1=N_1(q_2,p_2)-M_1(q_1,p_1),\qquad {\cal L}'_2=N_2(q_3,p_3)-M_2(q_2,p_2).
\]
Here,
\begin{gather*}
M_j=\frac12\int \frac{f_j(q_j)\, dq_j}{\sqrt{{\cal L}_j-V_j(q_j)-f_j(q_j){\cal L}_{j+1}}},\\
 N_j=\frac12\int \frac{dq_{j+1}}{\sqrt{{\cal L}_{j+1}-V_{j+1}(q_{j+1})-f_{j+1}(q_{j+1}){\cal L}_{j+2}}},\qquad j=1,2,
\end{gather*}
and ${\cal L}_4\equiv 0$. With this construction the  functions ${\cal L}_1$, ${\cal L}_2$, ${\cal L}_3$, ${\cal L}'_1$, ${\cal L}'_2$
are functionally independent constants of the motion. The functions ${\cal L}_2$, ${\cal L}_3$ are second order polynomials in the momenta and determine the separation of variables in
coordinates $q_1$, $q_2$, $q_3$. In general the functions ${\cal L}'_1$, ${\cal L}'_2$ are only locally def\/ined and are not polynomials. The system will be superintegrable only if we can
supplement ${\cal L}_1$, ${\cal L}_2$, ${\cal L}_3$ with two more polynomial functions such that the full set is functionally independent. This will be possible only for very special
systems  In the following we will look at several candidate systems for superintegrability,  show how to construct polynomial constants of the motion
from ${\cal L}'_1$, ${\cal L}'_2$ and
 work out the structure of the symmetry algebra generated by these constants.

\section[The classical 3D extended Kepler-Coulomb system with 3-parameter potential]{The classical 3D extended Kepler--Coulomb system\\ with 3-parameter potential}\label{3d3k}

The extended Kepler--Coulomb Hamiltonian is
\begin{gather*}
{\cal H}=p_r^2+\frac{\alpha}{r}+\frac{{\cal L}_2}{r^2},
\end{gather*}
where
\begin{gather*}
{\cal L}_2 =p_{\theta_1}^2+\frac{{\cal L}_3}{\sin^2(k_1\theta_1)},\qquad
{\cal L}_3= p_{\theta_2}^2+\frac{\beta}{\cos^2(k_2\theta_2)}+\frac{\gamma}{\sin^2(k_2\theta_2)}.
\end{gather*}
Here, ${\cal L}_2$, ${\cal L}_3$ are constants of the motion that determine additive separation of the Hamilton--Jacobi equation. Further
$\{{\cal L}_2,{\cal L}_3\}=0$ so ${\cal L}_2$ and ${\cal L}_3$ are in involution.

Applying our action angle construction to get two independent constants of the motion  we note that $q_1=r$, $q_2=\theta_1$, $q_3=\theta_2$ and
\begin{gather*}
f_1=\frac{1}{r^2},\qquad f_2=\frac{1}{\sin^2(k_1\theta_1)},\qquad f_3=0,\\ V_1=\frac{\alpha}{r},\qquad
V_2=0,\qquad V_3=\frac{\beta}{\cos^2(k_2\theta_2)}+\frac{\gamma}{\sin^2(k_2\theta_2)},
\end{gather*}
to obtain functions $A_j$, $B_j$, $j=1,2$ such that
\begin{gather*}
\sinh {A_1}=-\frac{i\sqrt{{\cal L}_2}\cos(k_1\theta_1)}{\sqrt{{\cal L}_2-{\cal L}_3}},\qquad
\cosh{A_1 }=
\frac{\sin (k_1\theta_1 )p_{\theta_1} }{\sqrt{{\cal L}_2-{\cal L}_3}},\\
\sinh{A_2}=\frac{i({\cal L}_3\cos(2k_2\theta_2)+\gamma-\beta) }{\sqrt{(\beta-\gamma-{\cal L}_3)^2-4\gamma{\cal L}_3}},\qquad
\cosh{A_2 }= -
\frac{\sqrt{ {\cal L}_3}\sin (2k_2\theta_2 )p_{\theta_2} }{\sqrt{(\beta-\gamma-{\cal L}_3)^2-4\gamma{\cal L}_3}},\\
\sinh{ B_1}= - \frac{i(\alpha+2{\cal L}_2/r)}{ \sqrt{ \alpha^2+4{\cal H} {\cal  L}_2}},\qquad
\cosh{B_1 }= \frac{2\sqrt{ {\cal L}_2}p_r}{ \sqrt{ \alpha^2+4{\cal H} {\cal L}_2}},\\
\sinh{ B_2} = \frac{i(2{\cal L}_3\csc^2(k_1\theta_1)-{\cal L}_2-{\cal L}_3)}{ {\cal L}_3-{\cal L}_2},\qquad
\cosh{B_2 }= -\frac{2\sqrt{ {\cal L}_3}\cot(k_1\theta_1)p_{\theta_1}}{ {\cal L}_3-{\cal L}_2}.
\end{gather*}
Here, $k_1=p_1/q_1$, $k_2=p_2/q_2$ where $p_1$, $q_1$ are relatively prime positive integers and $p_2$, $q_2$ are relatively prime positive integers.

From  our general theory,
\[N_1=-\frac{iA_1}{2k_1\sqrt{{\cal L}_2}},\qquad M_1=-\frac{iB_1}{2\sqrt{{\cal L}_2}},\qquad N_2=-\frac{iA_2}{4k_2\sqrt{{\cal L}_3}},\qquad M_2=-\frac{iB_2}{4k_1\sqrt{{\cal L}_3}},\]
so
\[p_1q_2A_2-p_2q_1B_2,\qquad q_1A_1-p_1B_1\]
are two  constants of the motion such that the full set of f\/ive constants of the motion is functionally independent.

We work with the exponential functions, \cite{CG}, see also~\cite{Tsiganov2008a,Tsiganov2009}.
We have, for $j=1,2$,
\begin{gather*}
e^{A_j}=\cosh A_j +\sinh A_j= {X_j}/{U_j},\qquad e^{-A_j}=\cosh A_j -\sinh A_j= {\overline{X_j}}/{U_j},\\
e^{B_j}=\cosh B_j +\sinh B_j= {Y_j}/{S_j},\qquad e^{-B_j}=\cosh B_j -\sinh B_j= {\overline{Y_j}}/{S_j},
\end{gather*}
where
\begin{gather*}
X_1= \sin  (k_1\theta_1) p_{\theta_1}-i\sqrt{{\cal L}_2}\cos(k_1\theta_1),\qquad
{\overline X_1}=\sin (k_1\theta_1) p_{\theta_1}+i\sqrt{{\cal L}_2}\cos(k_1\theta_1),\\
X_2= -\sqrt{{\cal L}_3}\sin (2k_2\theta_2) p_{\theta_2}+i({\cal L}_3\cos(2k_2\theta_2)+\gamma-\beta),\\
{\overline X_2}=-\sqrt{{\cal L}_3}\sin (2k_2\theta_2) p_{\theta_2}-i({\cal L}_3\cos(2k_2\theta_2)+\gamma-\beta),\\
Y_1=2\sqrt{{\cal L}_2}p_r-i\left(\alpha+2\frac{{\cal L}_2}{r}\right),\qquad
{\overline Y_1}=2\sqrt{{\cal L}_2}p_r+i\left(\alpha+2\frac{{\cal L}_2}{r}\right),\\
Y_2= -2\sqrt{{\cal L}_3} \cot(k_1\theta_1)p_{\theta_1}+i\big(2{\cal L}_3\csc^2(k_1\theta_1)-{\cal L}_2-{\cal L}_3\big), \\
{\overline Y_2}=-2\sqrt{{\cal L}_3} \cot\big(k_1\theta_1)p_{\theta_1}-i(2{\cal L}_3\csc^2(k_1\theta_1)-{\cal L}_2-{\cal L}_3\big),\\
 U_1=\sqrt{{\cal L}_2-{\cal L}_3},\qquad U_2=\sqrt{-(\beta-\gamma-{\cal L}_3)^2+4\gamma {\cal L}_3},\\
S_1=\sqrt{\alpha^2+4{\cal H}{\cal L}_2}, \qquad S_2={\cal L}_3-{\cal L}_2.
\end{gather*}
(Here, $\overline X$, $\overline Y$ are, in general,  not the complex conjugates of $X$, $Y$, respectively, unless all of the coordinates are real.)

Now note that $e^{q_1A_1-p_1B_1}$ and $e^{-q_1A_1+p_1B_1}$ are constants of the motion, where
\begin{gather*}
e^{q_1A_1-p_1B_1}=\big(e^{A_1}\big)^{q_1} \big(e^{-B_1}\big)^{p_1}=\frac{X_1^q\overline{Y_1}^{p_1}}{U_1^{q_1}S_1^{p_1}},
\qquad e^{-q_1A_1+p_1B_1}=\big(e^{-A_1}\big)^{q_1}
\big(e^{B_1}\big)^{p_1}=\frac{\overline{X_1}^{q_1}
Y_1^{p_1}}{U_1^{q_1}S_1^{p_1}}.
\end{gather*}
Moreover, the identity $e^{q_1A_1-p_1B_1}e^{-q_1A_1+p_1B_1}=1$ can be expressed as
\begin{gather*}
X_1^{q_1}\overline{X_1}^{q_1}Y_1^{p_1}\overline{Y_1}^{p_1}=U_1^{2q_1}S_1^{2p_1}={\cal P}_1({\cal H},{\cal L}_2, {\cal L}_3)
=({\cal L}_2-{\cal L}_3)^{q_1}
(\alpha^2+4{\cal H}{\cal L}_2)^{p_1},
\end{gather*}
where
${\cal P}_1$ is a polynomial in ${\cal H}$, ${\cal L}_2$ and ${\cal L}_3$.

Similarly, $e^{p_1q_2A_2-p_2q_1B_2}$
and $e^{-p_1q_2A_2+p_2q_1B_2}$ are constants of the motion, where
\begin{gather*}
e^{p_1q_2A_2-p_2q_1B_2}=\big(e^{A_2}\big)^{p_1q_2} \big(e^{-B_1}\big)^{p_2q_1}
=\frac{X_2^{p_1q_2}\overline{Y_2}^{p_2q_1}}{U_2^{p_1q_2}S_2^{p_2q_1}},
\nonumber\\
e^{-p_1q_2A_2+p_2q_1B_2)}=\big(e^{-A_1}\big)^{p_1q_2}
\big(e^{B_1}\big)^{p_2q_1}=\frac{\overline{X_2}^{p_1q_2}
Y_2^{p_2q_1}}{U_2^{p_1q_2}S_2^{p_2q_1}}.
\end{gather*}
 The identity $e^{p_1q_2A_2-p_2q_1B_2}e^{-p_1q_2A_2+p_2q_1B_2}=1 $ can be written as
\begin{gather*}
X_2^{p_1q_2}\overline{X_2}^{p_1q_2}Y_2^{p_2q_1}\overline{Y_2}^{p_2q_1}
=U_2^{2p_1q_2}S_2^{2p_2q_1}={\cal P}_2({\cal H},{\cal L}_2, {\cal L}_3)\nonumber\\
\hphantom{X_2^{p_1q_2}\overline{X_2}^{p_1q_2}Y_2^{p_2q_1}\overline{Y_2}^{p_2q_1}}{}
=({\cal L}_2-{\cal L}_3)^{2p_2q_1}
((\beta-\gamma-{\cal L}_3)^2-4\gamma{\cal L}_3)^{p_1q_2},
\end{gather*}
where ${\cal P}_2$ is a polynomial in ${\cal H}$, ${\cal L}_2$ and ${\cal L}_3$.

 Let $a$, $b$, $c$, $d$ be nonzero complex numbers and consider the
binomial expansion
\begin{gather}
\big(\sqrt{{\cal L}_k}a+ib\big)^q\big(\sqrt{{\cal L}_k}c+id\big)^p+\big(\sqrt{{\cal L}_k}a-ib\big)^q\big(\sqrt{{\cal L}_k}c-id\big)^p
\nonumber\\
\qquad{}
=\sum_{0\le \ell\le q, 0\le s\le p} \begin{pmatrix} q\\ \ell\end{pmatrix}
 \begin{pmatrix} p\\ s\end{pmatrix} b^\ell d^s a^{q-\ell}c^{p-s}{\cal L}_k^{(q+p-\ell-s)/2}\big[i^{\ell+s}+(-i)^{\ell+s}\big].\label{plusbinom}
\end{gather}
Here, either $k=2$ or $k=3$ and $p,q$ are positive integers.
Suppose $p+q$ is odd. Then it is easy to see that the sum (\ref{plusbinom}) takes the form
 $\sqrt{{\cal L}_k}T_{\rm odd}({\cal L}_k) $ where
$T_{\rm odd}$ is a polynomial in ${\cal L}_k$. On the other hand, if $p+q$ is even then the sum (\ref{plusbinom}) takes the
form $T_{\rm even}({\cal L}_k) $ where
$T_{\rm even}$ is a~polynomial in ${\cal L}_k$.

Similarly, consider the
binomial expansion
\begin{gather}
\frac{1}{i}\left[\big(\sqrt{{\cal L}_k}a+ib\big)^q\big(\sqrt{{\cal L}_k}c+id\big)^p-
\big(\sqrt{{\cal L}_k}a-ib\big)^q\big(\sqrt{{\cal L}_k}c-id\big)^p\right]
\nonumber\\
\qquad{} =\sum_{0\le \ell\le q, 0\le s\le p} \begin{pmatrix} q\\ \ell\end{pmatrix}
 \begin{pmatrix} p\\ s\end{pmatrix} b^\ell d^s a^{q-\ell}c^{p-s}\frac{{\cal L}_k^{(q+p-\ell-s)/2}}{i}\big[i^{\ell+s}-(-i)^{\ell+s}\big].\label{minusbinom}
\end{gather}
Suppose $p+q$ is odd. Then  the sum (\ref{minusbinom}) takes the form
 $V_{\rm odd}({\cal L}_k) $ where
$V_{\rm odd}$ is a polynomial in~${\cal L}_k$. On the other hand, if $p+q$ is even then the sum~(\ref{minusbinom}) takes the
form $\sqrt{{\cal L}_k}V_{\rm even}({\cal L}_k) $ where
$V_{\rm even}$ is a polynomial in~${\cal L}_k$.

A third possibility is
\begin{gather*}
\big(a+i\sqrt{{\cal L}_k}b\big)^q\big(\sqrt{{\cal L}_k}c+id\big)^p+\big(a-\sqrt{{\cal L}_k}b\big)^q\big(\sqrt{{\cal L}_k}c-id\big)^p
\nonumber\\
\qquad{}
=\sum_{0\le \ell\le q, 0\le s\le p} \begin{pmatrix} q\\ \ell\end{pmatrix}
 \begin{pmatrix} p\\ s\end{pmatrix} b^\ell d^s a^{q-\ell}c^{p-s}{\cal L}_k^{(\ell+p-s)/2}\big[i^{\ell+s}+(-i)^{\ell+s}\big].
\end{gather*}
Then we must have $p$ even to get a polynomial in ${\cal L}_k$. If $p$ is odd the sum takes the form $\sqrt{{\cal L}_k}T({\cal L}_k)$ where $T$ is a polynomial.

A fourth possibility is
\begin{gather*}
\big(a+i\sqrt{{\cal L}_k}b\big)^q\big(\sqrt{{\cal L}_k}c+id\big)^p-\big(a-\sqrt{{\cal L}_k}b\big)^q\big(\sqrt{{\cal L}_k}c-id\big)^p
\nonumber\\
\qquad{}=\sum_{0\le \ell\le q, 0\le s\le p} \begin{pmatrix} q\\ \ell\end{pmatrix}
 \begin{pmatrix} p\\ s\end{pmatrix} b^\ell d^s a^{q-\ell}c^{p-s}{\cal L}_k^{(\ell+p-s)/2}[i^{\ell+s}-(-i)^{\ell+s}].
\end{gather*}
Then we must have $p$ odd to get a polynomial in ${\cal L}_k$. If $p$ is even the sum takes the form $\sqrt{{\cal L}_k}T({\cal L}_k)$ where $T$ is a polynomial.

We def\/ine basic raising and lowering symmetries
\[
{\cal J}^+=X_1^{q_1}\overline{Y_1}^{p_1},\qquad {\cal J}^-=\overline{X_1}^{q_1}Y_1^{p_1},
\qquad  {\cal K}^+=X_2^{p_1q_2}\overline{Y_2}^{p_2q_1},\qquad {\cal K}^-
=\overline{X_2}^{p_1q_2}Y_2^{p_2q_1}.
\]
At this point we restrict to the case where each of $p_1$, $q_1$, $p_2$, $q_2$ is an odd integer. (The other cases are very similar.)

Let
\begin{gather*} 
{\cal J}_1=\frac{1}{\sqrt{{\cal L}_2}}({\cal J}^- +{\cal J}^+),\qquad
{\cal J}_2=\frac{1}{i}({\cal J}^- -{\cal J}^+),
\\
 {\cal K}_1=\frac{1}{i\sqrt{{\cal L}_3}}({\cal K}^- -{\cal K}^+),\qquad
{\cal K}_2={\cal K}^- +{\cal K}^+.
\end{gather*}

Then we see from the explicit expressions for the symmetries and the preceding parity argument  that ${\cal J}_1$, ${\cal J}_2$, ${\cal K}_1$, ${\cal K}_2$ are constants of the motion,
 polynomial in the momenta. Moreover, the identities
 \[{\cal J}^+{\cal J}^-={\cal P}_1,\qquad  {\cal K}^+{\cal K}^-={\cal P}_2\]
 hold.

 The following relations are straightforward to derive from the def\/inition of the Poisson bracket:
 \begin{gather*}
  \{{\cal L}_3,X_1\}= \{{\cal L}_3,\overline{X_1}\}=\{{\cal L}_3,Y_1\}= \{{\cal L}_3,\overline{Y_1}\}=0,\\
 \{{\cal L}_2,Y_1\}= \{{\cal L}_2,\overline{Y_1}\}=\{{\cal L}_3,Y_2\}= \{{\cal L}_3,\overline{Y_2}\}=0,\\
 \{{\cal L}_2,X_1\}= -2ik_1\sqrt{{\cal L}_2}X_1,\qquad \{{\cal L}_2,\overline{X_1}\}= 2ik_1\sqrt{{\cal L}_2}\overline{X_1},\\
 \{{\cal L}_3,X_2\}=- 4ik_2\sqrt{{\cal L}_3}X_2,\qquad \{{\cal L}_3,\overline{X_2}\}= 4ik_2\sqrt{{\cal L}_3}\overline{X_2},\\
\{{\cal L}_2,X_2\}= -\frac{4ik_2}{\sin^2(k_1\theta_1)}\sqrt{{\cal L}_3}X_2,\qquad \{{\cal L}_2,\overline{X_2}\}= \frac{4ik_2\sqrt{{\cal L}_3}}{\sin^2(k_1\theta_1)}
 \overline{X_2}, \\
 \{{\cal L}_2,Y_2\}= -\frac{4ik_1\sqrt{{\cal L}_3}}{\sin^2(k_1\theta_1)}Y_2,\qquad \{{\cal L}_2,\overline{Y_2}\}= \frac{4ik_1\sqrt{{\cal L}_3}}{\sin^2(k_1\theta_1)}
   \overline{Y_2},\\
\{{\cal H},X_1\}=\frac{4ik_1\sqrt{{\cal L}_2}}{r^2}X_1,\qquad \{{\cal H},\overline{X_1}\}=-\frac{4ik_1\sqrt{{\cal L}_2}}{r^2}\overline{X_1},\\
\{{\cal H},X_2\}=\frac{4ik_2\sqrt{{\cal L}_3}}{r^2\sin^2(k_1\theta_1)}X_2,\qquad \{{\cal H},\overline{X_2}\}=
-\frac{4ik_2\sqrt{{\cal L}_3}}{r^2\sin^2(k_1\theta_1)}\overline{X_2},\\
\{{\cal H},Y_1\}=\frac{2i\sqrt{{\cal L}_2}}{r^2}Y_1,\qquad \{{\cal H},\overline{Y_1}\}=-\frac{2i\sqrt{{\cal L}_2}}{r^2}\overline{Y_1},\\
\{{\cal H},Y_2\}=\frac{4ik_1\sqrt{{\cal L}_3}}{r^2\sin^2(k_1\theta_1)}Y_2,\qquad \{{\cal H},\overline{Y_2}\}=
-\frac{4ik_1\sqrt{{\cal L}_3}}{r^2\sin^2(k_1\theta_1)}\overline{Y_2},
\end{gather*}

 From these results, we f\/ind
\begin{gather*}
\{{\cal L}_3,{\cal J}^\pm\}=0,\qquad \{{\cal L}_2,{\cal J}^\pm\}=\mp 2ip_1\sqrt{{\cal L}_2}{\cal J}^\pm,\\
 \{{\cal L}_2,{\cal K}^\pm\}=0,\qquad \{{\cal L}_3,{\cal K}^\pm\}=\mp 4ip_1p_2\sqrt{{\cal L}_3}{\cal K}^\pm.
\end{gather*}

Thus  we obtain
\begin{gather*}
\{{\cal L}_2,{\cal J}_2\}=-i\big(2ip_1\sqrt{{\cal L}_2}{\cal J}^-+2ip_1\sqrt{{\cal L}_2}{\cal J}^+\big)=2p_1{\cal L}_2{\cal J}_1,\\
 \{{\cal L}_2,{\cal J}_1\}=\frac{1}{\sqrt{{\cal L}_2}}\big(2ip_1\sqrt{{\cal L}_2}{\cal J}^--2ip_1\sqrt{{\cal L}_2}{\cal L}^+\big)=-2p_1{\cal J}_2,\qquad
 \{{\cal L}_3,{\cal J}_1\}=\{{\cal L}_3,{\cal J}_2\}=0.
 \end{gather*}
Similarly,
\begin{gather*}
\{{\cal L}_3,{\cal K}_2\}=-4p_1p_2{\cal L}_3{\cal K}_1,\qquad \{{\cal L}_3,{\cal K}_1\}=4p_1p_2{\cal K}_2,\qquad
 \{{\cal L}_2,{\cal K}_2\}=\{{\cal L}_2,{\cal K}_1\}=0.
\end{gather*}

Since
\begin{gather}\label{L4L4}
{\cal J}_1^2=\frac{1}{{\cal L}_2}\left[ ({\cal J}^+)^2+2{\cal J}^+{\cal J}^-+({\cal J}^-)^2\right],\qquad
{\cal J}_2^2=-\left[ ({\cal J}^+)^2-2{\cal J}^+{\cal J}^-+({\cal J}^-)^2\right],
\end{gather}
 we have
$ {\cal J}_2^2+{\cal L}_2{\cal J}_1^2=4{\cal J}^+{\cal J}^-=4{\cal P}_1$, so
\begin{gather*}
{\cal J}_2^2=-{\cal L}_2{\cal J}_1^2+4{\cal P}_1({{\cal H}, \cal L}_2,{\cal L}_3).
\end{gather*}

Further,
\begin{gather*}
\{{\cal J}^+,{\cal J}^-\} =\left\{{\cal J}^+, \frac{{\cal P}_1}{{\cal J}^+}\right\}
=  ({\cal J}^+)^{-1}\{{\cal J}^+,{\cal P}_1\}\\
\hphantom{\{{\cal J}^+,{\cal J}^-\}}{}
=({\cal J}^+)^{-1}\left(\frac{\partial {\cal P}_1}{\partial {\cal L}_2}\{{\cal J}^+,{\cal L}_2\}+\frac{\partial {\cal P}_1}{\partial {\cal L}_3}\{{\cal J}^+,{\cal L}_3\}\right),
\end{gather*}
so
\begin{gather*}
\{{\cal J}^+,{\cal J}^-\}=2ip_1\sqrt{{\cal L}_2}\frac{\partial {\cal P}_1}{\partial{\cal L}_2}.
\end{gather*}

To evaluate $\{{\cal J}_2,{\cal J}_1\}$ we have
\begin{gather*}
\{{\cal J}_2,{\cal J}_1\}=\frac{1}{i}\{{\cal J}^--{\cal J}^+,\frac{{\cal J}^++{\cal J}^-}{\sqrt{{\cal L}_2}}\}\\
\hphantom{\{{\cal J}_2,{\cal J}_1\}}{}
= \frac{1}{i}\left[-\frac12 ({\cal L}_2)^{-3/2}({\cal J}^-+{\cal J}^+)\{{\cal L}_2,{\cal J}^+-{\cal J}^-\}
 +({\cal L}_2)^{-1/2}\{{\cal J}^--{\cal J}^+,{\cal J}^-+{\cal J}^+\}\right]\\
\hphantom{\{{\cal J}_2,{\cal J}_1\}}{}
 =-\frac{p_1}{{\cal L}_2}({\cal J}^-+{\cal J}^+)^2-4p_1\frac{\partial {\cal P}_1}{\partial{\cal L}_2}.
\end{gather*}
Then, using (\ref{L4L4}), we conclude that
\begin{gather*}
\{{\cal J}_2,{\cal J}_1\}=-p_1{\cal J}_1^2-4p_1\frac{\partial {\cal P}_1}{\partial{\cal L}_2}.
\end{gather*}

Similarly, we have the ${\cal K}$-related identities
\begin{gather*}
{\cal K}_1^2=-\frac{1}{{\cal L}_3}\left[ ({\cal K}^+)^2-2{\cal K}^+{\cal K}^-+({\cal K}^-)^2\right],\\
{\cal K}_2^2=\left[ ({\cal K}^+)^2+2{\cal K}^+{\cal K}^-+({\cal K}^-)^2\right],\qquad
{\cal K}_2^2=-{\cal L}_3{\cal K}_1^2+4{\cal P}_2({{\cal H}, \cal L}_2,{\cal L}_3),\\
\{{\cal K}^+,{\cal K}^-\}=4ip_1p_2\sqrt{{\cal L}_3}\frac{\partial {\cal P}_2}{\partial{\cal L}_3},\qquad
\{{\cal K}_2,{\cal K}_1\}=-2p_1p_2{\cal K}_1^2+8p_1p_2\frac{\partial {\cal P}_2}{\partial{\cal L}_3}.
\end{gather*}

Commutators relating the $\cal J$ and $\cal K$ symmetries are somewhat more complicated to compute. We have
\begin{gather*}
\{X_1,X_2\}=-\frac{2k_2\sqrt{{\cal L}_3}\cot^2(k_1\theta_1)}{\sin(k_1\theta_1)\sqrt{{\cal L}_2}}X_2,\\
 \{X_1,\overline{Y_2}\}= \frac{2k_1\sqrt{{\cal L}_3}\cot(k_1\theta_1)}{\sqrt{{\cal L}_2}\sin^2(k_1\theta_1)}\overline{Y_2}-2k_1\sqrt{{\cal L}_2}X_1
 +k_1\left(\cos(k_1\theta_1)p_{\theta_1}+i\sqrt{{\cal L}_2}\sin(k_1\theta_1)\right)\\
 \hphantom{\{X_1,\overline{Y_2}\}=}{}
 \times \big(2\sqrt{{\cal L}_3}\cot(k_1\theta_1)\big)-k_1\sin(k_1\theta_1)\left(\frac{2\sqrt{{\cal L}_3}p_{\theta_1}}
{\sin^2(k_1\theta_1)}+\frac{4i{\cal L}_3\cos(k_1\theta_1)}{\sin^3(k_1\theta_1)}\right),\\
  \{X_2,\overline{Y_1}\} =
\left(\frac{4ip_r}{\sqrt{{\cal L}_2}}-\frac{8}{r}\right)\frac{k_2\sqrt{{\cal L}_3}}{\sin^2(k_1\theta_1)}X_2.
\end{gather*}

Now we can determine the nonpolynomial constant of the motion $\{{\cal J}^+,{\cal K}^+\}$:
\begin{gather*}
\{{\cal J}^+,{\cal K}^+\}=\left\{X_1^{q_1}\overline{Y_1^{p_1}},X_2^{p_1q_2}\overline{Y_2^{p_2q_1}}\right\}\\
\hphantom{\{{\cal J}^+,{\cal K}^+\}}{}
 =q_1X_1^{q_1-1}\overline{Y_1^{p_1}}\left\{X_1,X_2^{p_1q_2}\overline{Y_2^{p_2q_1}}\right\}
 +p_1X_1^{q_1}\overline{Y_1^{p_1-1}}\left\{\overline{Y_1},X_2^{p_1q_2}
\overline{Y_2^{p_2q_1}}\right\}\\
\hphantom{\{{\cal J}^+,{\cal K}^+\}}{}
=q_1X_1^{q_1-1}\overline{Y_1^{p_1}}\left(p_1q_2X_2^{p_1q_2-1}\overline{Y_2^{p_2q_1}}\{X_1,X_2\}+p_2q_1X_2^{p_1q_2}\overline{Y_2^{p_2q_1-1}}\{X_1,
 \overline{Y_2}\}\right)\\
\hphantom{\{{\cal J}^+,{\cal K}^+\}=}{}
+p_1X_1^{q_1}\overline{Y_1^{p_1-1}}\left(p_2q_1X_2^{p_1q_2}\overline{Y_2^{p_2q_1-1}}\{\overline{Y_1},\overline{Y_2}\}
+p_1q_2X_2^{p_1q_2-1}\overline{Y_2^{p_2q_1}}\{\overline{Y_1},X_2\}\right),
\end{gather*}
where the last term in braces vanishes identically. We conclude that
\[
\frac{ \{ {\cal J}^+,{\cal K}^+ \}}{{\cal J}^+{\cal K}^+}
=\frac{2iq_1p_1p_2}{{\cal L}_2-{\cal L}_3}(\sqrt{{\cal L}_2}+\sqrt{{\cal L}_3}).
\]

Once we have $\{{\cal J}^+,{\cal K}^+\}$ explicitly, we can obtain the remaining Poisson relations between the ${\cal J}$ and $\cal K$ symmetries with
little additional work. We use the fact that ${\cal J}^+{\cal J}^-={\cal P}_1$ and ${\cal K}^+{\cal K}^-={\cal P}_2$. Then we have
\begin{gather*}
\{{\cal J}^-,{\cal K}^-\}=\left\{\frac{{\cal P}_1}{{\cal J}^+},\frac{{\cal P}_2}{{\cal K}^+}\right\}=-\frac{{\cal P}_1}{({\cal J}^+)^2}\left\{{\cal J}^+,\frac{{\cal P}_2}{{\cal K}^+}\right\} +\frac{1}{{\cal J}^+}\left\{{\cal P}_1,\frac{{\cal P}_2}{{\cal K}^+}\right\}\\
\hphantom{\{{\cal J}^-,{\cal K}^-\}}{}
=-\frac{{\cal P}_1}{({\cal J}^+)^2{\cal K}^+}\{{\cal J}^+,{\cal P}_2\}+\frac{{\cal P}_1{\cal P}_2}{({\cal J}^+)^2({\cal K}^+)^2}\{{\cal J}^+,{\cal K}^+\}
-\frac{{\cal P}_2}{{\cal J}^+({\cal K}^+)^2})\{{\cal J}^+,{\cal K}^+\}\\
\hphantom{\{{\cal J}^-,{\cal K}^-\}}{}
=-\frac{{\cal P}_1}{({\cal J}^+)^2{\cal K}^+}\frac{\partial{\cal P}_2}{\partial {\cal L}_2}\{{\cal J}^+\!,{\cal L}_2\}\!+\frac{{\cal P}_1{\cal P}_2}{({\cal J}^+)^2({\cal K}^+)^2}\{{\cal J}^+\!,{\cal K}^+\}\!-\frac{{\cal P}_2}{{\cal J}^+({\cal K}^+)^2}\frac{\partial{\cal P}_1}{\partial {\cal L}_3}\{{\cal L}_3,{\cal K}^+\}\\
\hphantom{\{{\cal J}^-,{\cal K}^-\}}{}
=-\frac{2ip_1\sqrt{{\cal L}_2}{\cal P}_1}{{\cal J}^+{\cal K}^+}\frac{\partial{\cal P}_2}{\partial {\cal L}_2}+
\frac{{\cal P}_1{\cal P}_2}{({\cal J}^+)^2({\cal K}^+)^2}\{{\cal J}^+,{\cal K}^+\}+\frac{4ip_1p_2\sqrt{{\cal L}_3}{\cal P}_2}{{\cal J}^+{\cal K}^+}
\frac{\partial{\cal P}_1}{\partial {\cal L}_3}.
\end{gather*}
We can write this relation in the more compact form
\[
\frac{\{{\cal J}^-,{\cal K}^-\}}{{\cal J}^-{\cal K}^-}=-4iq_1p_1p_2\frac{\sqrt{{\cal L}_2}+\sqrt{{\cal L}_3}}{{\cal L}_2-{\cal L}_3}+\frac{\{{\cal J}^+,{\cal K}^+\}}{{\cal J}^+{\cal K}^+}.
\]
Similarly, we have
\begin{gather*}
\frac{\{{\cal J}^+,{\cal K}^-\}}{{\cal J}^+{\cal K}^-}=4iq_1p_1p_2\frac{\sqrt{{\cal L}_2}}{{\cal L}_2-{\cal L}_3}-\frac{\{{\cal J}^+,{\cal K}^+\}}{{\cal J}^+{\cal K}^+},\\
\frac{\{{\cal J}^-,{\cal K}^+\}}{{\cal J}^-{\cal K}^+}=4iq_1p_1p_2\frac{\sqrt{{\cal L}_3}}{{\cal L}_2-{\cal L}_3}-\frac{\{{\cal J}^+,{\cal K}^+\}}{{\cal J}^+{\cal K}^+}.
\end{gather*}
Set $\{{\cal J}^+,{\cal K}^+\}={\cal Q}{\cal J}^+{\cal K}^+$. Then
\begin{gather*}
\{{\cal J}_1,{\cal K}_1\}=-i\left\{\frac{{\cal J}^-+{\cal J}^+}{\sqrt{{\cal L}_2}},\frac{{\cal K}^--{\cal K}^+}{\sqrt{{\cal L}_3}}\right\}=\frac{-i}{\sqrt{{\cal L}_2{\cal L}_3}}
\{{\cal J}^-+{\cal J}^+,{\cal K}^--{\cal K}^+\}\\
\hphantom{\{{\cal J}_1,{\cal K}_1\}}{}
 =\frac{-i}{\sqrt{{\cal L}_2{\cal L}_3}}\bigg(\frac{-4ip_1p_2q_1}{{\cal L}_2-{\cal L}_3}\left[\sqrt{{\cal L}_2}({\cal J}^--{\cal J}^+){\cal K}^-
+\sqrt{{\cal L}_3}({\cal K}^-+{\cal K}^+){\cal J}^-\right]\\
\hphantom{\{{\cal J}_1,{\cal K}_1\}=}{}
+({\cal J}^--{\cal J}^+)({\cal K}^-+{\cal K}^+){\cal Q}\bigg),
\end{gather*}
 where
\[
{\cal Q}=\frac{2iq_1p_1p_2}{{\cal L}_2-{\cal L}_3}\big(\sqrt{{\cal L}_2}+\sqrt{{\cal L}_3}\big).
\]
Thus,
\begin{gather*}
\{{\cal J}_1,{\cal K}_1\}=\frac{2q_1p_1p_2}{\sqrt{{\cal L}_2{\cal L}_3}({\cal L}_2-{\cal L}_3)}\\
\hphantom{\{{\cal J}_1,{\cal K}_1\}=}{}
\times\left[-
\big(\sqrt{{\cal L}_2}+\sqrt{{\cal L}_3}\big)\big({\cal J}^-{\cal K}^-+{\cal J}^+{\cal K}^+\big)+
\big(\sqrt{{\cal L}_2}-\sqrt{{\cal L}_3}\big)\big({\cal J}^-{\cal K}^+ +{\cal J}^+{\cal K}^-\big)\right].
\end{gather*}

In summary:
 \begin{gather*}
 {\cal J}^+{\cal J}^-={\cal P}_1,\qquad  {\cal K}^+{\cal K}^-={\cal P}_2,\qquad
 {\cal P}_1({\cal H},{\cal L}_2, {\cal L}_3)=({\cal L}_2-{\cal L}_3)^{2q_1}
(\alpha^2+4{\cal H}{\cal L}_2)^{p_1},\\
 {\cal P}_2({\cal H},{\cal L}_2, {\cal L}_3)
=({\cal L}_2-{\cal L}_3)^{2p_2q_1}
((\beta-\gamma-{\cal L}_3)^2-4\gamma{\cal L}_3)^{p_1q_2},\\
 \{{\cal L}_3,{\cal J}^\pm\}=0,\qquad \{{\cal L}_2,{\cal J}^\pm\}=\mp 2ip_1\sqrt{{\cal L}_2}{\cal J}^\pm,\\
 \{{\cal L}_2,{\cal K}^\pm\}=0,\qquad \{{\cal L}_3,{\cal K}^\pm\}=\mp 4ip_1p_2\sqrt{{\cal L}_3}{\cal K}^\pm,\\
 \{{\cal J}^+,{\cal J}^-\}=2ip_1\sqrt{{\cal L}_2}\frac{\partial {\cal P}_1}{\partial{\cal L}_2},\qquad
  \{{\cal K}^+,{\cal K}^-\}=4ip_1p_2\sqrt{{\cal L}_3}\frac{\partial {\cal P}_2}{\partial{\cal L}_3},\\
 \frac{\{{\cal J}^+,{\cal K}^-\}}{{\cal J}^+{\cal K}^-}=-\frac{\{{\cal J}^-,{\cal K}^+\}}{{\cal J}^-{\cal K}^+}=\frac{2iq_1p_1p_2(\sqrt{{\cal L}_2}-
\sqrt{{\cal L}_3})}{{\cal L}_2-{\cal L}_3},\\
 \frac{\{{\cal J}^-,{\cal K}^-\}}{{\cal J}^-{\cal K}^-}=- \frac{\{{\cal J}^+,{\cal K}^+\}}{{\cal J}^+{\cal K}^+}=
-\frac{2iq_1p_1p_2(\sqrt{{\cal L}_3}+\sqrt{{\cal L}_2})}{{\cal L}_2-{\cal L}_3}.
\end{gather*}
These relations prove closure of the symmetry algebra in the space of functions polynomial in~${\cal J}^\pm$,~${\cal K}^\pm$, rational
 in ${\cal L}_2$, ${\cal L}_3$, ${\cal H}$ and at most linear in $\sqrt{{\cal L}_2}$, $\sqrt{{\cal L}_3}$.

\subsection{Structure relations for polynomial constants of the motion}
Since,
\begin{gather*}
{\cal J}^-=\frac12(\sqrt{{\cal L}_2}{\cal J}_1+i{\cal J}_2),\qquad {\cal J}^+=\frac12(\sqrt{{\cal L}_2}{\cal J}_1-i{\cal J}_2),\\
 {\cal K}^-=\frac12(i\sqrt{{\cal L}_3}{\cal K}_1+{\cal K}_2),\qquad {\cal K}^+=\frac12(-i\sqrt{{\cal L}_3}{\cal K}_1+{\cal K}_2),
 \end{gather*}
we have
\[\{{\cal J}_1,{\cal K}_1\}=\frac{2q_1p_1p_2}{{\cal L}_2-{\cal L}_3}(-{\cal J}_1{\cal K}_2+{\cal J}_2{\cal K}_1).\]
Similar computations yield
\begin{gather*}
\{{\cal J}_2,{\cal K}_1\}=\frac{2q_1p_1p_2}{{\cal L}_2-{\cal L}_3}({\cal J}_2{\cal K}_2+{\cal L}_2{\cal J}_1{\cal K}_1),\qquad
 \{{\cal J}_1,{\cal K}_2\}=\frac{2q_1p_1p_2}{{\cal L}_2-{\cal L}_3}({\cal L}_3{\cal K}_1+{\cal J}_2{\cal K}_2),\\
 \{{\cal J}_2,{\cal K}_2\}=\frac{2q_1p_1p_2}{{\cal L}_2-{\cal L}_3}({\cal L}_3{\cal J}_2{\cal K}_1-{\cal L}_2{\cal J}_1{\cal K}_2).
\end{gather*}
Note: It can be verif\/ied that the numerators are divisible by ${\cal L}_2-{\cal L}_3$, so that $\{{\cal J}_1,{\cal K}_1\}$, $\{{\cal J}_2,{\cal K}_1\}$
$\{{\cal J}_1,{\cal K}_2\}$ and $\{{\cal J}_2,{\cal K}_2\}$ are true polynomial constants of the motion, although not polynomial in the generators.

\subsection{Minimal order generators} \label{mog3}
The generators for the polynomial symmetry algebra that we have produced so far are not of minimal order. Note that
 ${\cal L}_1$, ${\cal L}_2$, ${\cal L}_3$ are of order 2 and the orders of ${\cal J}_1$, ${\cal K}_1$ are one less than
the orders of ${\cal J}_2$, ${\cal K}_2$, respectively. We will construct a  symmetry ${\cal K}_0$ of order one less
than ${\cal K}_1$.
Note that the symmetry ${\cal K}_2$ is a polynomial in ${\cal L}_3$. The constant term
in this polynomial expansion is $i^{p_1q_2+p_2q_1}{\cal L}_2^{p_2q_1}(\gamma-\beta)^{p_1q_2}((-1)^{p_1q_2}+(-1)^{p_2q_1})$, itself a constant of the motion.
In the case we are considering $p_1$, $q_1$, $p_2$, $q_2$ are each odd, so the constant term is
\[
{\cal D}_2({\cal L}_2)=2(-1)^{(p_1q_2+p_2q_1)/2+1}{\cal L}_2^{p_2q_1}
(\gamma-\beta)^{p_1q_2}.
\]
 Thus
\[
{\cal K}_0=\frac{{\cal K}_2-{\cal D}_2}{{\cal L}_3}
\]
is a polynomial symmetry of order two less than ${\cal K}_2$. We have the identity
\begin{gather}\label{K2K0ident}
{\cal K}_2={\cal L}_3{\cal K}_0+{\cal D}_2.
\end{gather}
From this,  $\{{\cal L}_3,{\cal K}_2\}={\cal L}_3 \{{\cal L}_3,{\cal K}_0\}$.
We already know that $\{{\cal L}_3,{\cal K}_2\}=-4p_1p_2{\cal L}_3{\cal K}_1$
so
\[ \{{\cal L}_3,{\cal K}_0\}=-4p_1p_2{\cal K}_1,\qquad  \{{\cal L}_2,{\cal K}_0\}=0.\]

The same construction fails for ${\cal J}_2$. It is a polynomial in ${\cal L}_2$, but the constant term in the expansion is not a constant of the motion.
Indeed, in the special case $k_1=k_2=1$,
the symmetry~${\cal J}_1$ is of minimal order~2, so ${\cal J}_1$ cannot be realized as a commutator.

Now we choose ${\cal L}_1$, ${\cal L}_2$, ${\cal L}_3$, ${\cal J}_1$, ${\cal K}_0$ as the generators of our algebra. We def\/ine the basic nonzero commutators as
 \[{\cal R}_1=\{{\cal L}_2,{\cal J}_1\}=-2p_1{\cal J}_2,
 \qquad {\cal R}_2=\{{\cal L}_3,{\cal K}_0\}=-4p_1p_2{\cal K}_1,
 \qquad {\cal R}_3=\{{\cal J}_1,{\cal K}_0\}.\]
Then we have
\[
\frac{{\cal R}_1^2}{4p_1^2}={\cal J}_2^2=-{\cal L}_2{\cal J}_1^2+4{\cal P}_1,
\]
a polynomial in the generators.
 Further,
\[ \frac{{\cal R}_2^2}{16p_1^2p_2^2}={\cal K}_1^2=\frac{-({\cal L}_3{\cal K}_0+{\cal D}_2)^2+4{\cal P}_1}{{\cal L}_3}
\]
which again can be verif\/ied to be a polynomial in the generators.
 Note, however, that ${\cal R}_1{\cal R}_2=8p_1^2p_2{\cal J}_2{\cal K}_1$, a product of Poisson brackets of the generators,  is not a polynomial in the generators,
although $({\cal R}_1{\cal R}_2)^2$ is such  a polynomial. Using the identity~(\ref{K2K0ident}) and the expression for $\{{\cal J}_1,{\cal K}_2\}$, it is
 easy to see that ${\cal R}_3$ is rationally related to~${\cal R}_2$. It is clear that all additional commutators can be expressed as rational functions of the constants of the motion already computed.

 We conclude that the polynomial symmetry algebra generated by the 5 basic generators and their 3 commutators closes rationally, but not polynomially.

\section[The classical 3D extended Kepler-Coulomb system with 4-parameter potential]{The classical 3D extended Kepler--Coulomb system\\ with 4-parameter potential}
\label{3d4k}

Now we consider the Hamiltonian
\[{\cal H}=p_r^2+\frac{\alpha}{r}+\frac{{\cal L}_2}{r^2},
\]
where
\[{\cal L}_2=p^2_{\theta_1}+\frac{{\cal L}_3}{\sin^2(k_1\theta_1)}+\frac{\delta}{\cos^2(k_1\theta_1)},\qquad {\cal L}_3=p^2_{\theta_2}+\frac{\beta}{\cos^2(k_2\theta_2)}
+\frac{\gamma}{\sin^2(k_2\theta_2)}.\]
Here ${\cal L}_2$, ${\cal L}_3$ are constants of the motion, in involution. They determine additive separation in the variables $r$, $\theta$, $\phi$.

Applying our usual construction to get two independent constants of the motion  we note that $q_1=r$, $q_2=\theta_1$, $q_3=\theta_2$ and
\begin{gather*}
f_1=\frac{1}{r^2},\qquad f_2=\frac{1}{\sin^2(k_1\theta_1)},\qquad f_3=0,\\
V_1=\frac{\alpha}{r},\qquad
V_2=\frac{\delta}{\cos^2(k_1\theta_1)},\qquad V_3=\frac{\beta}{\cos^2(k_2\theta_2)}+\frac{\gamma}{\sin^2(k_2\theta_2)},
\end{gather*}
to obtain functions $A_j$, $B_j$, $j=1,2$ such that
\begin{gather*}
\sinh {A_1}=\frac{i(-{\cal L}_2\cos(2k_1\theta_1)+\delta-{\cal L}_3)}{\sqrt{{\cal L}_3^2-2{\cal L}_3({\cal L}_2+\delta)+({\cal L}_2-\delta)^2}},\\
\cosh{A_1 }=
\frac{\sqrt{{\cal L}_2}\sin (2k_1\theta_1 )p_{\theta_1} }{\sqrt{{\cal L}_3^2-2{\cal L}_3({\cal L}_2+\delta)+({\cal L}_2-\delta)^2}},\\
\sinh{A_2}=\frac{i({\cal L}_3\cos(2k_2\theta_2)+\gamma-\beta) }{\sqrt{(\beta-\gamma-{\cal L}_3)^2-4\gamma{\cal L}_3}},\qquad
\cosh{A_2 }= -
\frac{\sqrt{ {\cal L}_3}\sin (2k_2\theta_2 )p_{\theta_2} }{\sqrt{(\beta-\gamma-{\cal L}_3)^2-4\gamma{\cal L}_3}},\\
\sinh{ B_1}= - \frac{i(\alpha+2{\cal L}_2/r)}{ \sqrt{ \alpha^2+4{\cal H} {\cal  L}_2}},\qquad
\cosh{B_1 }= \frac{2\sqrt{ {\cal L}_2}p_r}{ \sqrt{ \alpha^2+4{\cal H} {\cal L}_2}},\\
\sinh{ B_2} = \frac{-2i\sqrt{{\cal L}_3}\cot(k_1\theta_1)p_{\theta_1}}{\sqrt{{\cal L}_3^2-2{\cal L}_3(\delta+{\cal L}_2)+({\cal L}_2-\delta)^2}},\\
\cosh{B_2 }= \frac{ -2{\cal L}_3\cot^2(k_1\theta_1)+({\cal L}_2-{\cal L}_3-\delta)}{\sqrt{{\cal L}_3^2-2{\cal L}_3(\delta+{\cal L}_2)+({\cal L}_2-\delta)^2}}.
\end{gather*}
Here, $k_1=p_1/q_1$, $k_2=p_2/q_2$ where $p_1$, $q_1$ are relatively prime positive integers and $p_2$, $q_2$ are relatively prime positive integers.

From  our general theory,
\[N_1=-\frac{iA_1}{4k_1\sqrt{{\cal L}_2}},\qquad M_1=-\frac{iB_1}{2\sqrt{{\cal L}_2}},\qquad N_2=-\frac{iA_2}{4k_2\sqrt{{\cal L}_3}},\qquad M_2=-\frac{iB_2}{4k_1\sqrt{{\cal L}_3}},\]
so
\[p_1q_2A_2-p_2q_1B_2,\qquad q_1A_1-2p_1B_1\]
are two  constants of the motion such that the full set of f\/ive constants of the motion is functionally independent.
We have
\begin{gather*}
e^{A_j}=\cosh A_j +\sinh A_j= {X_j}/{U_j},\qquad e^{-A_j}=\cosh A_j -\sinh A_j= {\overline{X_j}}/{U_j},\\
  e^{B_j}=\cosh B_j +\sinh B_j= {Y_j}/{S_j},\qquad e^{-B_j}=\cosh B_j -\sinh B_j= {\overline{Y_j}}/{S_j},
 \end{gather*}
where
\begin{gather*}
 X_1= \sqrt{{\cal L}_2}\sin (2k_1\theta_1) p_{\theta_1}+i(-{\cal L}_2\cos(2k_1\theta_1)+\delta-{\cal L}_3),\\
 {\overline X_1}=\sqrt{{\cal L}_2}\sin (2k_1\theta_1) p_{\theta_1}-i(-{\cal L}_2\cos(2k_1\theta_1)+\delta-{\cal L}_3),\\
X_2= -\sqrt{{\cal L}_3}\sin (2k_2\theta_2) p_{\theta_2}+i({\cal L}_3\cos(2k_2\theta_2)+\gamma-\beta),\\
{\overline X_2}=-\sqrt{{\cal L}_3}\sin (2k_2\theta_2) p_{\theta_2}-i({\cal L}_3\cos(2k_2\theta_2)+\gamma-\beta),\\
Y_1=2\sqrt{{\cal L}_2}p_r-i\left(\alpha+2\frac{{\cal L}_2}{r}\right),\qquad
{\overline Y_1}=2\sqrt{{\cal L}_2}p_r+i\left(\alpha+2\frac{{\cal L}_2}{r}\right),\\
Y_2= -2{\cal L}_3 \cot^2(k_1\theta_1)+({\cal L}_2-{\cal L}_3-\delta)-2i\sqrt{{\cal L}_3}\cot(k_1\theta_1)p_{\theta_1}, \\
{\overline Y_2}=-2{\cal L}_3 \cot^2(k_1\theta_1)+({\cal L}_2-{\cal L}_3-\delta)+2i\sqrt{{\cal L}_3}\cot(k_1\theta_1)p_{\theta_1},\\
U_1=\sqrt{{\cal L}^2_3-2{\cal L}_3({\cal L}_2+\delta)+({\cal L}_2-\delta)^2},\qquad U_2=\sqrt{(\beta-\gamma-{\cal L}_3)^2-4\gamma {\cal L}_3},\\
S_1=\sqrt{\alpha^2+4{\cal H}{\cal L}_2}, \qquad S_2=\sqrt{{\cal L}^2_3-2{\cal L}_3({\cal L}_2+\delta)+({\cal L}_2-\delta)^2}.
\end{gather*}

Here, $e^{q_1A_1-2p_1B_1}$ and $e^{-q_1A_1+p_1B_1}$ are constants of the motion, where
\begin{gather*}
e^{q_1A_1-2p_1B_1}=\big(e^{A_1}\big)^{q_1} \big(e^{-B_1}\big)^{2p_1}=\frac{X_1^{q_1}\overline{Y_1}^{2p_1}}{U_1^{q_1}S_1^{2p_1}},\nonumber\\
 e^{-q_1A_1+2p_1B_1}=\big(e^{-A_1}\big)^{q_1}
\big(e^{B_1}\big)^{2p_1}=\frac{\overline{X_1}^{q_1}
Y_1^{2p_1}}{U_1^{q_1}S_1^{2p_1}}.
\end{gather*}
The identity $e^{q_1A_1-p_1B_1}e^{-q_1A_1+p_1B_1}=1$ can be expressed as
\begin{gather*}
X_1^{q_1}\overline{X_1}^{q_1}Y_1^{2p_1}\overline{Y_1}^{2p_1}=U_1^{2q_1}S_1^{4p_1}={\cal P}_1({\cal H},{\cal L}_2, {\cal L}_3)\nonumber\\
\hphantom{X_1^{q_1}\overline{X_1}^{q_1}Y_1^{2p_1}\overline{Y_1}^{2p_1}}{}
=\big[{\cal L}^2_3-2{\cal L}_3({\cal L}_2+\delta)+({\cal L}_2-\delta)^2\big]^{q_1}
\big[\alpha^2+4{\cal H}{\cal L}_2\big]^{2p_1},
\end{gather*}
where
${\cal P}_1$ is a polynomial in ${\cal H}$, ${\cal L}_2$ and ${\cal L}_3$.

Similarly, $e^{p_1q_2A_2-p_2q_1B_2}$,  $e^{-p_1q_2A_2+p_2q_1B_2}$ are constants of the motion, where
\begin{gather*}
e^{p_1q_2A_2-p_2q_1B_2}=\big(e^{A_2}\big)^{p_1q_2} \big(e^{-B_1}\big)^{p_2q_1}
=\frac{X_2^{p_1q_2}\overline{Y_2}^{p_2q_1}}{U_2^{p_1q_2}S_2^{p_2q_1}},
\nonumber\\
e^{-p_1q_2A_2+p_2q_1B_2)}=\big(e^{-A_1}\big)^{p_1q_2}
\big(e^{B_1}\big)^{p_2q_1}=\frac{\overline{X_2}^{p_1q_2}
Y_2^{p_2q_1}}{U_2^{p_1q_2}S_2^{p_2q_1}}.
\end{gather*}
 The identity $e^{p_1q_2A_2-p_2q_1B_2}e^{-p_1q_2A_2+p_2q_1B_2}=1 $ becomes
\begin{gather*}
X_2^{p_1q_2}\overline{X_2}^{p_1q_2}Y_2^{p_2q_1}\overline{Y_2}^{p_2q_1}
=U_2^{2p_1q_2}S_2^{2p_2q_1}={\cal P}_2({\cal H},{\cal L}_2, {\cal L}_3)\\ 
\hphantom{X_2^{p_1q_2}\overline{X_2}^{p_1q_2}Y_2^{p_2q_1}\overline{Y_2}^{p_2q_1}}{}
=\big[(\beta-\gamma-{\cal L}_3)^2-4\gamma {\cal L}_3\big]^{p_1q_2}
\big[{\cal L}^2_3-2{\cal L}_3({\cal L}_2+\delta)+({\cal L}_2-\delta)^2\big]^{p_2q_1},\nonumber
\end{gather*}
where ${\cal P}_2$ is a polynomial in ${\cal H}$, ${\cal L}_2$ and ${\cal L}_3$.

We def\/ine basic raising and lowering symmetries
\[ {\cal J}^+=X_1^{q_1}\overline{Y_1}^{2p_1},\qquad
 {\cal J}^-=\overline{X_1}^{q_1}Y_1^{2p_1},\qquad  {\cal K}^+=X_2^{p_1q_2}\overline{Y_2}^{p_2q_1},\qquad {\cal K}^-
=\overline{X_2}^{p_1q_2}Y_2^{p_2q_1}.\]
Further, we restrict to the case where each of $p_1$, $q_1$, $p_2$, $q_2$ is an odd integer. The other cases are similar.

Let
\begin{gather*} 
{\cal J}_1=\frac{1}{\sqrt{{\cal L}_2}}({\cal J}^- +{\cal J}^+),\qquad
{\cal J}_2=\frac{1}{i}({\cal J}^- -{\cal J}^+),\\
{\cal K}_1=\frac{1}{\sqrt{{\cal L}_3}}({\cal K}^- +{\cal K}^+),\qquad {\cal K}_2=\frac{1}{i}({\cal K}^- -{\cal K}^+).
\end{gather*}
Then we see from the explicit expressions for the symmetries that ${\cal J}_1$, ${\cal J}_2$, ${\cal K}_1$, ${\cal K}_2$ are constants of the motion,
 polynomial in the momenta. Moreover, the identities
 \[{\cal J}^+{\cal J}^-={\cal P}_1,\qquad  {\cal K}^+{\cal K}^-={\cal P}_2\]
 hold. Note that the def\/initions of the $\cal K$-symmetries dif\/fer from those for the 3-parameter potential.

 The following relations are straightforward to derive from the def\/inition of the Poisson bracket:
 \begin{gather*}
  \{{\cal L}_3,X_1\}= \{{\cal L}_3,\overline{X_1}\}=\{{\cal L}_3,Y_1\}= \{{\cal L}_3,\overline{Y_1}\}=0,\\
 \{{\cal L}_2,Y_1\}= \{{\cal L}_2,\overline{Y_1}\}=\{{\cal L}_3,Y_2\}= \{{\cal L}_3,\overline{Y_2}\}=0,\\
 \{{\cal L}_2,X_1\}= -4ik_1\sqrt{{\cal L}_2}X_1,\qquad \{{\cal L}_2,\overline{X_1}\}= 4ik_1\sqrt{{\cal L}_2}\overline{X_1},\\
 \{{\cal L}_3,X_2\}=- 4ik_2\sqrt{{\cal L}_3}X_2,\qquad \{{\cal L}_3,\overline{X_2}\}= 4ik_2\sqrt{{\cal L}_3}\overline{X_2},\\
\{{\cal L}_2,X_2\}= -\frac{4ik_2}{\sin^2(k_1\theta_1)}\sqrt{{\cal L}_3}X_2,\qquad \{{\cal L}_2,\overline{X_2}\}= \frac{4ik_2\sqrt{{\cal L}_3}}
{\sin^2(k_1\theta_1)}\overline{X_2}, \\
 \{{\cal L}_2,Y_2\}= -\frac{4ik_1\sqrt{{\cal L}_3}}{\sin^2(k_1\theta_1)}Y_2,\qquad \{{\cal L}_2,\overline{Y_2}\}=
\frac{4ik_1\sqrt{{\cal L}_3}}{\sin^2(k_1\theta_1)}\overline{Y_2},\\
\{{\cal H},X_1\}=-\frac{4ik_1\sqrt{{\cal L}_2}}{r^2}X_1,\qquad \{{\cal H},\overline{X_1}\}= \frac{4ik_1\sqrt{{\cal L}_2}}{r^2}\overline{X_1},\\
\{{\cal H},X_2\}= -\frac{4ik_2}{r^2\sin^2(k_1\theta_1)}\sqrt{{\cal L}_3}X_2,\qquad \{{\cal H},\overline{X_2}\}= \frac{4ik_2\sqrt{{\cal L}_3}}{r^2\sin^2(k_1\theta_1)}\overline{X_2}, \\
 \{{\cal H},Y_1\}= -\frac{2i\sqrt{{\cal L}_2}}{r^2}Y_1,\qquad  \{{\cal H},\overline{Y_1}\}= \frac{2i\sqrt{{\cal L}_2}}{r^2}\overline{Y_1},\\
 \{{\cal H},Y_2\}= -\frac{4ik_1\sqrt{{\cal L}_3}}{r^2\sin^2(k_1\theta_1)}Y_2,\qquad \{{\cal H},\overline{Y_2}\}= \frac{4ik_1\sqrt{{\cal L}_3}}{r^2
\sin^2(k_1\theta_1)}\overline{Y_2}.
\end{gather*}

From these results, we f\/ind
\begin{gather*}
\{{\cal L}_3,{\cal J}^\pm\}=0,\qquad \{{\cal L}_2,{\cal J}^\pm\}=\mp 4ip_1\sqrt{{\cal L}_2}{\cal J}^\pm,\\
\{{\cal L}_2,{\cal K}^\pm\}=0,\qquad \{{\cal L}_3,{\cal K}^\pm\}=\mp 4ip_1p_2\sqrt{{\cal L}_3}{\cal K}^\pm.
\end{gather*}

Further,
\begin{gather*}
\{{\cal J}^+,{\cal J}^-\} =\left\{{\cal J}^+, \frac{{\cal P}_1}{{\cal J}^+}\right\}
=  ({\cal J}^+)^{-1}\{{\cal J}^+,{\cal P}_1\}\\
\hphantom{\{{\cal J}^+,{\cal J}^-\}}{}
=({\cal J}^+)^{-1}
\left(\frac{\partial {\cal P}_1}{\partial {\cal L}_2}\{{\cal J}^+,{\cal L}_2\}+\frac{\partial {\cal P}_1}{\partial {\cal L}_3}\{{\cal J}^+,{\cal L}_3\}\right),
\end{gather*}
so
\begin{gather*}
\{{\cal J}^+,{\cal J}^-\}=4ip_1\sqrt{{\cal L}_2}\frac{\partial {\cal P}_1}{\partial{\cal L}_2}.
\end{gather*}
Similarly
\begin{gather*}
\{{\cal K}^+,{\cal K}^-\}=4ip_1p_2\sqrt{{\cal L}_3}\frac{\partial {\cal P}_2}{\partial{\cal L}_3}.
\end{gather*}

Commutators relating the $\cal J$ and $\cal K$ symmetries are somewhat more complicated to compute. We have
\begin{gather*}
\{X_1,X_2\}=-\frac{4k_2\sqrt{{\cal L}_3}}{\sqrt{{\cal L}_2}}\left(i\cot(k_1\theta_1)p_{\theta_1}+\sqrt{{\cal L}_2}\cot^2(k_1\theta_1)\right)X_2,\\
 \{\overline{Y_1},\overline{Y_2}\}=\frac{4ik_1\sqrt{{\cal L}_3}}{\sin^2(k_1\theta_1)}\left( \frac{p_r}{\sqrt{{\cal L}_2}}+\frac{2i}{r}\right)\overline{Y_2},\\
 \{X_1,\overline{Y_2}\}= 4ik_1\sqrt{{\cal L}_2}X_1+ \frac{4ik_1\sqrt{{\cal L}_3}}{\sin^2(k_1\theta_1)}\left(\frac{\sin(2k_1\theta_1)p_{\theta_1}}{2\sqrt{{\cal L}_2}}
-i\cos(2k_1\theta_1)\right)\overline{Y_2}\\
\hphantom{\{X_1,\overline{Y_2}\}=}{}
+8k_1\cot^2(k_1\theta_1)\left(\frac{i\sqrt{{\cal L}_2{\cal L}_3}\sin(2k_1\theta_1)}{2}p_{\theta_1}-
\sqrt{{\cal L}_3}{\cal L}_2\sin^2(k_1\theta_1)-\sqrt{{\cal L}_2}{\cal L}_3\right),\\
\{X_2,\overline{Y_1}\} =
\frac{4ik_2\sqrt{{\cal L}_3}}{\sin^2(k_1\theta_1)}\left(\frac{p_r}{\sqrt{{\cal L}_2}}+\frac{2i}{r}\right)X_2.
\end{gather*}

Now we can compute the nonpolynomial constant of the motion $\{{\cal J}^+,{\cal K}^+\}$:
\begin{gather*}
\{{\cal J}^+,{\cal K}^+\}=\left\{X_1^{q_1}\overline{Y_1^{2p_1}},X_2^{p_1q_2}\overline{Y_2^{p_2q_1}}\right\}\\
\hphantom{\{{\cal J}^+,{\cal K}^+\}}{}
 =q_1X_1^{q_1-1}\overline{Y_1^{2p_1}}\left\{X_1,X_2^{p_1q_2}\overline{Y_2^{p_2q_1}}\right\}
 +2p_1X_1^{q_1}\overline{Y_1^{2p_1-1}}\left\{\overline{Y_1},X_2^{p_1q_2}\overline{Y_2^{p_2q_1}}\right\}\\
\hphantom{\{{\cal J}^+,{\cal K}^+\}}{}
=q_1X_1^{q_1-1}\overline{Y_1^{2p_1}}\left(p_1q_2X_2^{p_1q_2-1}\overline{Y_2^{p_2q_1}}\{X_1,X_2\}
+p_2q_1X_2^{p_1q_2}\overline{Y_2^{p_2q_1-1}}\{X_1,\overline{Y_2}\}\right)\\
\hphantom{\{{\cal J}^+,{\cal K}^+\}=}{}
+2p_1X_1^{q_1}\overline{Y_1^{2p_1-1}}\left(p_2q_1X_2^{p_1q_2}\overline{Y_2^{p_2q_1-1}}\{\overline{Y_1},\overline{Y_2}\}
+p_1q_2X_2^{p_1q_2-1}\overline{Y_2^{p_2q_1}}\{\overline{Y_1},X_2\}\right),
\end{gather*}
where the last term in braces vanishes identically. We conclude that
\[
\frac{ \{ {\cal J}^+,{\cal K}^+ \}}{{\cal J}^+{\cal K}^+}
=\frac{4iq_1p_1p_2(\sqrt{{\cal L}_2}-\sqrt{{\cal L}_3})({\cal L}_2+2\sqrt{{\cal L}_2{\cal L}_3}+{\cal L}_3
-\delta)}
{({\cal L}_3-{\cal L}_2-\delta)^2-
4\delta{\cal L}_2}.
\]

Once we have $\{{\cal J}^+,{\cal K}^+\}$ explicitly, we can obtain the remaining Poisson relations between the ${\cal J}$ and $\cal K$ symmetries with
little additional work. We use the fact that ${\cal J}^+{\cal J}^-={\cal P}_1$ and ${\cal K}^+{\cal K}^-={\cal P}_2$. Then we have
\begin{gather*}
\{{\cal J}^-,{\cal K}^-\}=\left\{\frac{{\cal P}_1}{{\cal J}^+},\frac{{\cal P}_2}{{\cal K}^+}\right\}
=-\frac{{\cal P}_1}{({\cal J}^+)^2}\left\{{\cal J}^+,\frac{{\cal P}_2}{{\cal K}^+}\right\}
+\frac{1}{{\cal J}^+}\left\{{\cal P}_1,\frac{{\cal P}_2}{{\cal K}^+}\right\}\\
\phantom{\{{\cal J}^-,{\cal K}^-\}}{}
=-\frac{{\cal P}_1}{({\cal J}^+)^2{\cal K}^+}\{{\cal J}^+,{\cal P}_2\}+\frac{{\cal P}_1{\cal P}_2}{({\cal J}^+)^2({\cal K}^+)^2}\{{\cal J}^+,{\cal K}^+\}
-\frac{{\cal P}_2}{{\cal J}^+({\cal K}^+)^2})\{{\cal P}_1,{\cal K}^+\}\\
\phantom{\{{\cal J}^-,{\cal K}^-\}}{}
=-\frac{{\cal P}_1}{({\cal J}^+)^2{\cal K}^+}\frac{\partial{\cal P}_2}{\partial {\cal L}_2}\{{\cal J}^+\!,{\cal L}_2\}\!+\frac{{\cal P}_1{\cal P}_2}{({\cal J}^+)^2({\cal K}^+)^2}\{{\cal J}^+\!,{\cal K}^+\}\!-\frac{{\cal P}_2}{{\cal J}^+({\cal K}^+)^2}\frac{\partial{\cal P}_1}{\partial {\cal L}_3}\{{\cal L}_3,{\cal K}^+\}\\
\phantom{\{{\cal J}^-,{\cal K}^-\}}{}
=-\frac{4ip_1\sqrt{{\cal L}_2}{\cal P}_1}{{\cal J}^+{\cal K}^+}\frac{\partial{\cal P}_2}{\partial {\cal L}_2}+
\frac{{\cal P}_1{\cal P}_2}{({\cal J}^+)^2({\cal K}^+)^2}\{{\cal J}^+,{\cal K}^+\}+\frac{4ip_1p_2\sqrt{{\cal L}_3}{\cal P}_2}{{\cal J}^+{\cal K}^+}
\frac{\partial{\cal P}_1}{\partial {\cal L}_3}.
\end{gather*}
We can write this relation in the more compact form
\[\frac{\{{\cal J}^-,{\cal K}^-\}}{{\cal J}^-{\cal K}^-}=8iq_1p_1p_2\frac{(\sqrt{{\cal L}_3}-\sqrt{{\cal L}_2})
({\cal L}_2+2\sqrt{{\cal L}_2{\cal L}_3}+{\cal L}_3-\delta)}{({\cal L}_3-{\cal L}_2-\delta)^2-
4\delta{\cal L}_2}+\frac{\{{\cal J}^+,{\cal K}^+\}}{{\cal J}^+{\cal K}^+}.
\]
Similarly, we have
\begin{gather*}
\frac{\{{\cal J}^+,{\cal K}^-\}}{{\cal J}^+{\cal K}^-}=8iq_1p_1p_2\frac{\sqrt{{\cal L}_2}(-{\cal L}_3+{\cal L}_2-\delta)}{({\cal L}_3-{\cal L}_2-\delta)^2-
4\delta{\cal L}_2}-\frac{\{{\cal J}^+,{\cal K}^+\}}{{\cal J}^+{\cal K}^+},\\
\frac{\{{\cal J}^-,{\cal K}^+\}}{{\cal J}^-{\cal K}^+}=
8iq_1p_1p_2\frac{\sqrt{{\cal L}_3}(-{\cal L}_3+{\cal L}_2+\delta)}{({\cal L}_3-{\cal L}_2-\delta)^2-
4\delta{\cal L}_2}-\frac{\{{\cal J}^+,{\cal K}^+\}}{{\cal J}^+{\cal K}^+}.
\end{gather*}
Thus,
\begin{gather*}
\frac{\{{\cal J}^+,{\cal K}^+\}}{{\cal J}^+{\cal K}^+}=-\frac{\{{\cal J}^-,{\cal K}^-\}}{{\cal J}^-{\cal K}^-}=
\frac{4iq_1p_1p_2(\sqrt{{\cal L}_2}-\sqrt{{\cal L}_3})({\cal L}_2+2\sqrt{{\cal L}_2{\cal L}_3}+{\cal L}_3
-\delta)}{({\cal L}_3-{\cal L}_2-\delta)^2-4\delta{\cal L}_2},\\
\frac{\{{\cal J}^+,{\cal K}^-\}}{{\cal J}^+{\cal K}^-}=-\frac{\{{\cal J}^-,{\cal K}^+\}}{{\cal J}^-{\cal K}^+}
=\frac{4iq_1p_1p_2(\sqrt{{\cal L}_2}+\sqrt{{\cal L}_3}) ({\cal L}_2-2\sqrt{{\cal L}_2{\cal L}_3}+{\cal L}_3
-\delta)    }{({\cal L}_3-{\cal L}_2-\delta)^2-4\delta{\cal L}_2}.
\end{gather*}

In summary:
 \begin{gather*}
 {\cal J}^+{\cal J}^-={\cal P}_1,\qquad  {\cal K}^+{\cal K}^-={\cal P}_2,\\
{\cal P}_1({\cal H},{\cal L}_2, {\cal L}_3)=\big[{\cal L}^2_3-2{\cal L}_3({\cal L}_2+\delta)+({\cal L}_2-\delta)^2\big]^{q_1}
\big[\alpha^2+4{\cal H}{\cal L}_2\big]^{2p_1},\\
{\cal P}_2({\cal H},{\cal L}_2, {\cal L}_3)
=\big[(\beta-\gamma-{\cal L}_3)^2-4\gamma {\cal L}_3\big]^{p_1q_2}
\big[{\cal L}^2_3-2{\cal L}_3({\cal L}_2+\delta)+({\cal L}_2-\delta)^2\big]^{p_2q_1},\\
\{{\cal L}_3,{\cal J}^\pm\}=0,\qquad \{{\cal L}_2,{\cal J}^\pm\}=\mp 4ip_1\sqrt{{\cal L}_2}{\cal J}^\pm,\\
\{{\cal L}_2,{\cal K}^\pm\}=0,\qquad \{{\cal L}_3,{\cal K}^\pm\}=\mp 4ip_1p_2\sqrt{{\cal L}_3}{\cal K}^\pm,\\
\{{\cal J}^+,{\cal J}^-\}=4ip_1\sqrt{{\cal L}_2}\frac{\partial {\cal P}_1}{\partial{\cal L}_2},\qquad
 \{{\cal K}^+,{\cal K}^-\}=4ip_1p_2\sqrt{{\cal L}_3}\frac{\partial {\cal P}_2}{\partial{\cal L}_3},\\
\frac{\{{\cal J}^+,{\cal K}^+\}}{{\cal J}^+{\cal K}^+}=-\frac{\{{\cal J}^-,{\cal K}^-\}}{{\cal J}^-{\cal K}^-}=
\frac{4iq_1p_1p_2(\sqrt{{\cal L}_2}-\sqrt{{\cal L}_3})({\cal L}_2+2\sqrt{{\cal L}_2{\cal L}_3}+{\cal L}_3
-\delta)}{({\cal L}_3-{\cal L}_2-\delta)^2-4\delta{\cal L}_2},\\
 \frac{\{{\cal J}^+,{\cal K}^-\}}{{\cal J}^+{\cal K}^-}=-\frac{\{{\cal J}^-,{\cal K}^+\}}{{\cal J}^-{\cal K}^+}
=\frac{4iq_1p_1p_2(\sqrt{{\cal L}_2}+\sqrt{{\cal L}_3}) ({\cal L}_2-2\sqrt{{\cal L}_2{\cal L}_3}+{\cal L}_3
-\delta)    }{({\cal L}_3-{\cal L}_2-\delta)^2-4\delta{\cal L}_2}.
\end{gather*}

These relations prove closure of the symmetry algebra in the space of functions polynomial in~${\cal J}^\pm$,~${\cal K}^\pm$, rational
 in ${\cal L}_2$, ${\cal L}_3$, ${\cal H}$ and at most linear in $\sqrt{{\cal L}_2}$, $\sqrt{{\cal L}_3}$.

 \subsection{Structure relations for polynomial symmetries\\ of the 4-parameter potential}\label{polystructure4}

Note that
\begin{gather*}
 {\cal J}^-=\frac12\big(\sqrt{{\cal L}_2}{\cal J}_1+i{\cal J}_2\big),\qquad  {\cal J}^+=\frac12\big(\sqrt{{\cal L}_2}{\cal J}_1-i{\cal J}_2\big),\\
{\cal K}^-=\frac12\big(i{\cal K}_2+\sqrt{{\cal L}_3}{\cal K}_1\big),\qquad  {\cal K}^+=\frac12\big({-}i{\cal K}_2+\sqrt{{\cal L}_3}{\cal K}_1\big).
\end{gather*}

Thus we have
\begin{gather}
\{{\cal J}_1,{\cal K}_1\}
=\left\{\frac{{\cal J}^-+{\cal J}^+}{\sqrt{{\cal L}_2}},\frac{{\cal K}^-+{\cal K}^+}{\sqrt{{\cal L}_3}}\right\}
=\frac{1}{\sqrt{{\cal L}_2{\cal L}_3}}
\{{\cal J}^-+{\cal J}^+,{\cal K}^-+{\cal K}^+\}\nonumber\\
\phantom{\{{\cal J}_1,{\cal K}_1\}}{} =\frac{4q_1p_1p_2}{{({\cal L}_3-{\cal L}_2-\delta)^2-4\delta{\cal L}_2}}
\left[{\cal J}_1{\cal K}_2({\cal L}_2-{\cal L}_3+\delta)+{\cal J}_2{\cal K}_1({\cal L}_2
-{\cal L}_3-\delta)\right].\label{ident1}
\end{gather}
Similarly,
\begin{gather}\{{\cal J}_1,{\cal K}_2\}=
 -\frac{4q_1p_1p_2}{({\cal L}_3\!-{\cal L}_2\!-\delta)^2\!-4\delta{\cal L}_2}\left[{\cal J}_1{\cal K}_1{\cal L}_3({\cal L}_2-{\cal L}_3+\delta)+{\cal J}_2{\cal K}_2
(-{\cal L}_2+{\cal L}_3+\delta)\right],\!\!\!\!\label{ident2}\\
\{{\cal J}_2,{\cal K}_2\}=
 -\frac{4q_1p_1p_2}{({\cal L}_3\!-{\cal L}_2\!-\delta)^2\!-4\delta{\cal L}_2}\left[{\cal J}_1{\cal K}_2{\cal L}_2({\cal L}_2-{\cal L}_3-\delta)+{\cal J}_2{\cal K}_1
{\cal L}_3({\cal L}_2
-{\cal L}_3+\delta)\right], \!\!\!\!\label{ident3}\\
\{{\cal J}_2,{\cal K}_1\}=
 -\frac{4q_1p_1p_2}{({\cal L}_3\!-{\cal L}_2\!-\delta)^2\!-4\delta{\cal L}_2}\left[{\cal J}_1{\cal K}_1{\cal L}_2({\cal L}_2-{\cal L}_3-\delta)+{\cal J}_2{\cal K}_2
(-{\cal L}_2+{\cal L}_3-\delta)\right]. \!\!\!\!\label{ident4}
\end{gather}

Straightforward computations yield
\begin{gather*}
\{{\cal L}_2,{\cal J}_2\}=4p_1{\cal L}_2{\cal J}_1,\qquad \{{\cal L}_2,{\cal J}_1\}=-4p_1{\cal J}_2,\qquad
 \{{\cal L}_3,{\cal J}_1\}=\{{\cal L}_3,{\cal J}_2\}=0,\\
 \{{\cal L}_3,{\cal K}_2\}=4p_1p_2{\cal L}_3{\cal K}_1,\qquad \{{\cal L}_3,{\cal K}_1\}=-4p_1p_2{\cal K}_2,\qquad
 \{{\cal L}_2,{\cal K}_2\}=\{{\cal L}_2,{\cal K}_1\}=0,\\
  {\cal J}_2^2=-{\cal L}_2{\cal J}_1^2+4{\cal P}_1({{\cal H}, \cal L}_2,{\cal L}_3),
\qquad {\cal K}_2^2=-{\cal L}_3{\cal K}_1^2+4{\cal P}_2({{\cal H}, \cal L}_2,{\cal L}_3),\\
 \{{\cal J}_1,{\cal J}_2\}=-2p_1{\cal J}_1^2+8p_1\frac{\partial {\cal P}_1}{\partial{\cal L}_2},\qquad
 \{{\cal K}_1,{\cal K}_2\}=-2p_1p_2{\cal K}_1^2+8p_1p_2\frac{\partial {\cal P}_2}{\partial{\cal L}_3}.
 \end{gather*}

\section{Minimal order generators}\label{mog4}

The generators for the polynomial symmetry algebra that we have produced so far are not of minimal order. Here
 ${\cal L}_1$, ${\cal L}_2$, ${\cal L}_3$ are of order 2 and the orders of ${\cal J}_1$, ${\cal K}_1$ are one less than
the orders of~${\cal J}_2$,~${\cal K}_2$, respectively. We will construct  symmetries~${\cal J}_0$,~${\cal K}_0$ of order one less
than~${\cal J}_1$,~${\cal K}_1$, respectively. (In the standard case $k_1=k_2=1$ it is easy to see that~${\cal J}_1$ is of order 5 and~${\cal J}_2$ is of order~6,
whereas ${\cal K}_1$ is of order~3 and ${\cal K}_2$ is of order~4. Then ${\cal J}_0$, ${\cal K}_0$  will be of orders~4 and~2, respectively, which we know
corresponds to the minimal generators of the symmetry algebra in this case,~\cite{DASK2011, Evans2008a}.)
The symmetry ${\cal J}_2$ is a polynomial in ${\cal L}_2$ with constant term
\[{\cal D}_1=2(-1)^{(q_1-1)/2}(\delta-{\cal L}_3)^{q_1}\alpha^{2p_1},\]
 itself a constant of the motion.
 Thus
\[ {\cal J}_0=\frac{{\cal J}_2-{\cal D}_1}{{\cal L}_2}\]
is a polynomial symmetry of order two less than ${\cal J}_2$. We have the identity
\[ {\cal J}_2={\cal L}_2{\cal J}_0+{\cal D}_1.\]
{}From this,  $\{{\cal L}_2,{\cal J}_2\}={\cal L}_2 \{{\cal L}_2,{\cal J}_0\}$. We already know that $\{{\cal L}_2,{\cal J}_2\}=4p_1{\cal L}_2{\cal J}_1$
so
\[ \{{\cal L}_2,{\cal J}_0\}=4p_1{\cal J}_1,\qquad  \{{\cal L}_3,{\cal J}_0\}=0.\]

The same construction works for ${\cal K}_2$. It is a polynomial in ${\cal L}_3$, with constant term
\[{\cal D}_2=2(-1)^{(p_1q_2+1)/2} (\gamma-\beta)^{p_1q_2}({\cal L}_2-\delta)^{p_2q_1}.\]
Thus
\[
{\cal K}_0=\frac{{\cal K}_2-{\cal D}_2}{{\cal L}_3}
\]
is a polynomial symmetry of order two less than ${\cal K}_2$ and  we have the identity
\[
{\cal K}_2={\cal L}_3{\cal K}_0+{\cal D}_2.
\]
Further,
\[  \{{\cal L}_3,{\cal K}_0\}=4p_1p_2{\cal K}_1,\qquad  \{{\cal L}_2,{\cal K}_0\}=0.\]

Now we choose ${\cal L}_1$, ${\cal L}_2$, ${\cal L}_3$, ${\cal J}_0$, ${\cal K}_0$ as the generators of our algebra. We def\/ine the basic nonzero commutators as
 \[
 {\cal R}_1=\{{\cal L}_2,{\cal J}_0\}=4p_1{\cal J}_1,\qquad {\cal R}_2=\{{\cal L}_3,{\cal K}_0\}=4p_1p_2{\cal K}_1, \qquad {\cal R}_3=\{{\cal J}_0,{\cal K}_0\}.
 \]

Then we have
\begin{gather}\label{R1^2}
 \frac{{\cal R}_1^2}{16p_1^2}={\cal J}_1^2=-{\cal L}_2{\cal J}_0^2-2{\cal D}_1{\cal J}_0+\frac{4{\cal P}_1-{\cal D}_1^2}{{\cal L}_2},
\end{gather}
where the last term on the right is a polynomial in the generators ${\cal H}$, ${\cal L}_2$, ${\cal L}_3$.
 Further,
\begin{gather}\label{R2^2}
\frac{{\cal R}_2^2}{16p_1^2p_2^2}={\cal K}_1^2=-{\cal L}_3{\cal K}_0^2-2{\cal D}_2{\cal K}_0+\frac{4{\cal P}_2-{\cal D}_2^2}{{\cal L}_3},
\end{gather}
which again can be verif\/ied to be a polynomial in the generators.
Note that this symmetry algebra cannot close polynomially in the usual sense. If it did close then the product ${\cal R}_1{\cal R}_2$
would be expressible as a polynomial in the generators. The preceding two equations show that ${\cal R}_1^2{\cal R}_2^2$  is so expressible, but that the resulting polynomial is
not a perfect square. Thus the only possibility to obtain closure is to add new generators to the algebra, necessarily functionally dependent on the original set.

We see that \begin{gather}
 \big[({\cal L}_3-{\cal L}_2-\delta)^2-4\delta{\cal L}_2\big]{\cal R}_3\nonumber\\
 \qquad{}=  -\frac{4q_1p_1p_2}{{\cal L}_2{\cal L}_3 }\left[
 {\cal J}_1({\cal L}_3{\cal K}_0+{\cal D}_2){\cal L}_2({\cal L}_2-{\cal L}_3-\delta)+({\cal L}_2{\cal J}_0+{\cal D}_1){\cal K}_1{\cal L}_3({\cal L}_2-{\cal L}_3+\delta)
\right]\nonumber\\
\qquad\quad{}
  + \frac{4p_1}{{\cal L}_2{\cal L}_3 }[({\cal L}_3-{\cal L}_2-\delta)^2-4\delta{\cal L}_2]\left[{\cal L}_2{\cal J}_1\frac{\partial {\cal D}_2}{\partial {\cal L}_2}-p_2{\cal L}_3{\cal K}_1
\frac{\partial {\cal D}_1}{\partial {\cal L}_3}\right]
 =A{\cal J}_1+B{\cal K}_1,\label{R3}
\end{gather}
where $A$ and $B$ are polynomial in the generators, so
\begin{gather}
\big[({\cal L}_3-{\cal L}_2-\delta)^2-4\delta{\cal L}_2\big]\{{\cal L}_2,{\cal R}_3\}\nonumber\\
\qquad{} =4p_1A({\cal L}_2{\cal J}_0+{\cal D}_1)
-16q_1p_1^2p_2({\cal L}_2-{\cal L}_3+\delta){\cal J}_1{\cal K}_1,\label{L2R3}\\
 \big[({\cal L}_3-{\cal L}_2-\delta)^2-4\delta{\cal L}_2\big]\{{\cal L}_3,{\cal R}_3\}\nonumber\\
 \qquad{} =-4p_1p_2B({\cal L}_3{\cal K}_0+{\cal D}_2)
-16q_1p_1^2p_2^2({\cal L}_2-{\cal L}_3-\delta){\cal J}_1{\cal K}_1. \label{L3R3}
\end{gather}
Similarly
\begin{gather}\label{L2R1}
\frac{1}{4p_1}\{{\cal L}_2,{\cal R}_1\}=\{{\cal L}_2,{\cal J}_1\}=-4p_1{\cal J}_2=4p_1({\cal L}_2{\cal J}_0+{\cal D}_1),
\qquad
 \frac{1}{4p_1}\{{\cal L}_3,{\cal R}_1\}=0,
 \end{gather}
and
\begin{gather}\label{J0R1} \frac{1}{4p_1}\{{\cal J}_0,{\cal R}_1\}=\{{\cal J}_0,{\cal J}_1\}=\frac{1}{{\cal L}_2}
\left(2p_1{\cal J}_1^2-8p_1\frac{\partial {\cal P}_1}{\partial {\cal L}_2}\right)
+2p_1\frac{\partial}{\partial {\cal L}_2}\left(\frac{{\cal D}_1}{{\cal L}_2}\right){\cal J}_2+4p_1\frac{{\cal J}_2^2}{{\cal L}_2^2},
\end{gather}
a polynomial in the generators,
\begin{gather}
\frac{1}{4p_1}\{{\cal K}_0,{\cal R}_1\}=\{{\cal K}_0,{\cal J}_1\}=
 \frac{4q_1p_1p_2}{{\cal L}_3[({\cal L}_3-{\cal L}_2-\delta)^2-4\delta{\cal L}_2]}\nonumber\\
 \hphantom{\frac{1}{4p_1}\{{\cal K}_0,{\cal R}_1\}=}{}
 \times\left({\cal J}_1
{\cal K}_1{\cal L}_3({\cal L}_2-{\cal L}_3+\delta)+{\cal J}_2
{\cal K}_2(-{\cal L}_2+{\cal L}_3+\delta)\right)
+4p_1\frac{{\cal J}_2}{{\cal L}_3}\frac{\partial {\cal D}_2}{\partial {\cal L}_2}.\label{K0R1}
\end{gather}

Continuing in this way, it is straightforward to show that ${\cal L}_1$, ${\cal L}_2$, ${\cal L}_3$, ${\cal K}_0$, ${\cal J}_0$ generate a~symmetry algebra that closes rationally.
In particular, each of the commutators~${\cal R}_1$,~${\cal R}_2$,~${\cal R}_3$ satisf\/ies an explicit polynomial equation in the generators.

\subsection[St\"ackel equivalence of Kepler-Coulomb and caged isotropic oscillator systems]{St\"ackel equivalence of Kepler--Coulomb\\ and caged isotropic oscillator systems}\label{stackel}

Consider the Hamiltonian  for the caged isotropic oscillator
\begin{gather}\label{isotropic}
{\cal H}'=p_R^2+\alpha'R^2+\frac{{\cal L}'_2}{R^2},
\end{gather}
where
\[
{\cal L}'_2=p^2_{\phi_1}+\frac{{\cal L}'_3}{\sin^2(j_1\phi_1)}+\frac{\delta'}{\cos^2(j_1\phi_1)},\qquad {\cal L}'_3=p^2_{\phi_2}+\frac{\beta'}{\cos^2(j_2\phi_2)}
+\frac{\gamma'}{\sin^2(j_2\phi_2)}.
\]
Here ${\cal L}'_2$, ${\cal L}'_3$ are constants of the motion, in involution. They determine additive separation in the spherical coordinates  $R$, $\phi_1$, $\phi_2$.
Also, $j_1$, $j_2$ are nonzero rational numbers. If $j_1=j_2=1$, then in terms of Cartesian coordinates
we have ${\cal H}'=p_x^2+p_y^2+p_z^2+ \alpha'R^2 +\beta'/x^2+\gamma'/y^2+\delta'/z^2$. Note that system~(\ref{isotropic})  can be considered as the
3-variable analog of the TTW system \cite{TTW1,TTW2}. (Note, however, that this is a f\/lat space system only if $j_1=1$.)  Now consider the Hamilton--Jacobi
equation ${\cal H}'=E'$ and take the St\"ackel transform that corresponds to dividing by $R^2$. Then, making the change of variables
 $r=R^2$, $2\phi_1=\theta_1$, $2\phi_2=\theta_2$, we obtain the new Hamilton--Jacobi equation
${\cal H}=E$ where
\[
{\cal H}= p_r^2+\frac{\alpha}{r} +\frac{{\cal L}_2}{r^2}
\]
with
\begin{gather*}
{\cal L}_2=p^2_{\theta_1}+\frac{{\cal L}_3}{\sin^2(k_1\theta_1)}+\frac{\delta}{\cos^2(k_1\theta_1)},\qquad {\cal L}_3=p^2_{\theta_2}+\frac{\beta}{\cos^2(k_2\theta_2)}
+\frac{\gamma}{\sin^2(k_2\theta_2)},\\
 E=-\alpha'/4,\quad \alpha=-E'/4, \quad \beta=\beta'/4,\quad \gamma=\gamma'/4,\quad \delta=\delta'/4,\quad k_1=j_1/2,\quad k_2=j_2/2.
\end{gather*}
In other words, we obtain the extended Kepler--Coulomb system. Since the St\"ackel transform preserves the structure of the symmetry algebra of a
superintegrable system \cite{VE2008,KMP10, PW2010,SB2008}, all of our structure results apply to the caged  isotropic oscillator. Note, however, that the standard case $k_1=k_2=1$ for
 Kepler--Coulomb corresponds to  $j_1=j_2=2$ for the oscillator. Further, only for the cases $k_1=1$ and $j_1=1$ is the manifold f\/lat. The similar analysis in two
dimensions~\cite{PW2010} is always restricted to f\/lat space, but here the manifolds depend on $k_1$ and $j_1$.

\subsection[The special case $k_1=k_2=1$]{The special case $\boldsymbol{k_1=k_2=1}$}\label{k1k2}

In the case $k_1=k_2=1$  we are in Euclidean space and our system has additional symmetry. In terms of Cartesian coordinates
\[x=r\sin\theta_1\cos\theta_2,\qquad y=r\sin\theta_1\sin\theta_2,\qquad z=r\cos\theta_1,  \]
the Hamiltonian is
\begin{gather*}
{\cal H}=p_x^2+p_y^2+p_z^2+\frac{\alpha}{r}+\frac{\beta}{x^2}+\frac{\gamma}{y^2}+\frac{\delta}{z^2}.
\end{gather*}
Note that any permutation of the ordered pairs $(x,\beta)$, $(y,\gamma)$, $(z,\delta)$  leaves the Hamiltonian unchanged. This leads to
additional structure in the symmetry algebra.
The basic symmetries are
\begin{gather*}
{\cal L}_2=(xp_y-yp_x)^2+(yp_z-zp_y)^2+(zp_x-xp_z)^2\nonumber\\
\hphantom{{\cal L}_2=}{}
+\frac{\beta(x^2+y^2+z^2)}{x^2}+\frac{\gamma(x^2+y^2+z^2)}{y^2}+\frac{\delta(x^2+y^2+z^2)}{z^2},
\\
{\cal L}_3={\cal I}_{xy}=(xp_y-yp_x)^2+\frac{\beta(x^2+y^2)}{x^2}+\frac{\gamma(x^2+y^2)}{y^2}.
\end{gather*}
Note that the permutation symmetry of the Hamiltonian shows that ${\cal I}_{xz}$, ${\cal I}_{yz}$ are also constants of the motion, and that
\[
{\cal L}_2= {\cal I}_{xy}+{\cal I}_{xz}+{\cal I}_{yz}-(\beta+\gamma+\delta).
\]
The constant of the motion ${\cal K}_0$ is 2nd order:
\[
{\cal K}_0=4{\cal I}_{yz}+2{\cal L}_3-2({\cal L}_2+\beta+\gamma+\delta)=2({\cal I}_{yz}-{\cal I}_{xz}),
\]
and ${\cal J}_0$ is 4th order:
\begin{gather*}
{\cal J}_0=-16\left({\cal M}_3^2+\frac{\delta(xp_x+yp_y+zp_z)^2}{z^2}\right)+8{\cal H}({\cal I}_{xz}+{\cal I}_{yz}-\beta-\gamma-\delta)+2\alpha^2,
\end{gather*}
where
\begin{gather*}
{\cal M}_3=(yp_z-zp_y)p_y-(zp_x-xp_z)p_x-z\left(\frac{\alpha}{2r}+\frac{\beta}{x^2}+\frac{\gamma}{y^2}+\frac{\delta}{z^2}\right).
\end{gather*}
If $\delta=0$ then ${\cal M}_3$ is the analog of the 3rd component of the Laplace vector and is itself a constant of the motion.

 The symmetries ${\cal H}$, ${\cal L}_2$, ${\cal L}_3$, ${\cal J}_0$, ${\cal K}_0$ form a generating (rational) basis for the constants of the motion.
Under the transposition $(x,\beta)\leftrightarrow(z,\delta)$ this basis is mapped to an alternate basis ${\cal H}$, ${\cal L}'_2$, ${\cal L}'_3$, ${\cal J}'_0$, ${\cal K}'_0$
where
\begin{gather}
 {\cal L}'_2 = {\cal L}_2,\qquad  {\cal L}'_3=\frac14{\cal K}_0+\frac12{\cal L}_2-\frac12{\cal L}_3+\frac{\beta+\gamma+\delta}{2},\nonumber\\
  {\cal K}'_0=\frac12{\cal K}_0-{\cal L}_2+3{\cal L}_3-(\beta+\gamma+\delta),\qquad
{\cal R}_1'=\{{\cal L}_2,{\cal J}_0'\},\qquad {\cal R}_2'= \{{\cal L}_3',{\cal K}_0'\}=-\frac54{\cal R}_2,\nonumber\\
 {\cal R}_3'=\{{\cal J}_0',{\cal K}_0'\}=2{\cal R}_1'-2\{{\cal L}_3,{\cal J}_0'\}, \label{R'idents}
 \end{gather}
since $\{{\cal L}_3,{\cal J}_0'\}=0$.

All of the identities in Section~\ref{polystructure4} hold for the primed symmetries. It is easy to see that the ${\cal K}'$ symmetries are simple polynomials in the ${\cal L}$, ${\cal K}$ symmetries already constructed, e.g., ${\cal K}'_1=\frac14\{{\cal L}'_3,{\cal K}'_0\}=-\frac54{\cal K}_1$.
However, the ${\cal J}'$ symmetries are new. In particular,
\begin{gather*}
 {\cal J}'_0=-16\left({\cal M}_1^2+\frac{\beta(xp_x+yp_y+zp_z)^2}{x^2}\right)+8{\cal H}({\cal I}_{xy}+{\cal I}_{xz}-\beta-\gamma-\delta)+2\alpha^2,
 \end{gather*}
where
\begin{gather*}
{\cal M}_1=(yp_x-xp_y)p_y-(xp_z-zp_x)p_z-x\left(\frac{\alpha}{2r}+\frac{\beta}{x^2}+\frac{\gamma}{y^2}+\frac{\delta}{z^2}\right).
\end{gather*}

Note that the transposition $(y,\gamma)\leftrightarrow(z,\delta)$ does not lead to anything new. Indeed, under the symmetry we would obtain a constant of the motion
\[
{\cal J}_0''=-16\left({\cal M}_2^2+\frac{\gamma(xp_x+yp_y+zp_z)^2}{y^2}\right)+8{\cal H}({\cal I}_{xy}+{\cal I}_{yz}-\beta-\gamma-\delta)+2\alpha^2,
\]
where
\begin{gather*}
{\cal M}_2=(zp_y-yp_z)p_z-(yp_x-xp_y)p_x-y\left(\frac{\alpha}{2r}+\frac{\beta}{x^2}+\frac{\gamma}{y^2}+\frac{\delta}{z^2}\right).
\end{gather*}
but it is straightforward to check that
\begin{gather}\label{Jident}
{\cal J}_0+{\cal J}'_0+{\cal J}''_0 =2\alpha^2,
\end{gather}
so that the new constant depends linearly on the previous constants.

For further use, note that under the transposition symmetry $(x,\beta)\leftrightarrow(y,\gamma)$ the constants of the motion  ${\cal L}_2$, ${\cal L}_3$, ${\cal J}_0$, ${\cal J}_1$
are invariant, whereas ${\cal K}_0$, ${\cal K}_1$ change sign. The action on ${\cal J}_0'$ is more complicated. Under the transposition
${\cal J}'_0$ and ${\cal J}_0'' $ switch places. Thus from expression (\ref{Jident}) we see that
\begin{gather*}
{\cal J}'_0\longrightarrow {\cal J}_0''=2\alpha^2-{\cal J}_0-{\cal J}_0'.
\end{gather*}

In the paper \cite{DASK2011}, Tanoudis and Daskaloyannis show that the quantum symmetry algebra generated by the 6 functionally dependent symmetries
${\cal H}$, ${\cal L}_2$, ${\cal L}_3$, ${\cal J}_0$, ${\cal K}_0$ and ${\cal J}'_0$  closes polynomially, in the sense that all double commutators of the generators are
again expressible as polynomials in the generators, very strong evidence that the classical analog also closes polynomially. (However, they did not address the issue of determining the functional
indenpendence explicitly.)  Keys to understanding the polynomial closure are the rational structure equations (\ref{R3})--(\ref{K0R1}) and the terms
 ${\cal J}_1{\cal K}_1$ and
\[
{\cal Q}=({\cal L}_3-{\cal L}_2-\delta)^2-4\delta{\cal L}_2.
\]
If ${\cal J}_1{\cal K}_1$ can be expressed as a polynomial in the generators then, with this result substituted in the rational structure equations,
we will obtain polynomial structure equations. The requirement that the rational structure equations become polynomial is a strong restriction on the polynomial
\[
{\cal J}_1{\cal K}_1=P({\cal H}, {\cal L}_2,{\cal L}_3,{\cal J}_0,{\cal K}_0,{\cal J}'_0).
\] Let's focus on equation~(\ref{K0R1}). Note that the left hand side
of this equation is of order 6 in the momenta and  ${\cal J}_1{\cal K}_1$ is of order~8. What can we say about the polynomial $P$ in order that its substitution into~(\ref{K0R1}) turns the right hand side into a polynomial structure equation?

First note that the polynomial is not determined uniquely by this requirement. Indeed, if~$P_1$,~$P_2$ are two solutions their dif\/ference is  of the form $S{\cal Q}$ where
$S$ is a polynomial in the generators of order $\le 4$ in the momenta. Similarly, we can add such a $S{\cal Q}$ to any solution and get another solution.
 A straightforward computation using the polynomial and degree conditions alone yields the result
\begin{gather} {\cal J}_1{\cal K}_1=\frac12[{\cal L}_2+{\cal L}_3-\delta]{\cal J}_0{\cal K}_0+\alpha^2[{\cal L}_2-3{\cal L}_3
-\delta]{\cal K}_0\nonumber\\
 \hphantom{{\cal J}_1{\cal K}_1=}{}
 +(\beta-\gamma)[3{\cal L}_2-{\cal L}_3+\delta]{\cal J}_0+2\alpha^2(\gamma-\beta)[{\cal L}_2+{\cal L}_3-5\delta]+S{\cal Q},\label{J1K1}\\
\{{\cal K}_0,{\cal J}_1\}=\frac14\{{\cal K}_0,{\cal R}_1\}=-\frac14\{{\cal L}_2,{\cal R}_3\}\nonumber\\
\hphantom{\{{\cal K}_0,{\cal J}_1\}}{}
=-2[2(\gamma-\beta)+{\cal K}_0]\big[{\cal J}_0-2\alpha^2\big]+4({\cal L}_2-{\cal L}_3+\delta)S,\label{K0R11}
\end{gather}
where
\[S=c_1{\cal J}_0+c_2{\cal J}'_0+{\cal S}_1,\]
$c_1$ and $c_2\ne 0$ are parameters, and ${\cal S}_1({\cal H},{\cal L}_2,{\cal L}_3,{\cal K}_0)$  is a polynomial of order $\le 2$ in its arguments.
 Under the transposition  $(x,\beta)\leftrightarrow(y,\gamma)$ the left hand sides of equations~(\ref{J1K1}) and~(\ref{K0R11}) change sign.
 Since ${\cal Q}$ is invariant, $S$ must change sign. With all of these hints it is fairly easy to compute the exact result. It is
 \begin{gather*}
 S=-{\cal J}_0-2{\cal J}'_0+2\alpha^2,
 \end{gather*}
 in agreement with all our conditions.
 To determine the functional relationship between the 6~generators we can use the identity $F\equiv {\cal J}_1^2{\cal K}_1^2-({\cal J}_1{\cal K}_1)^2 =0$,
 where the f\/irst term is obtained from (\ref{R1^2}) and (\ref{R2^2}) and the second term  from  (\ref{J1K1}). The result is a polynomial in the momenta of order 16, and of the simple form
 \begin{gather*}
 F\equiv {\cal Q}\left({\cal A}_1({\cal J}_0')^2+{\cal A}_2{\cal J}_0'{\cal J}_0+{\cal A}_3({\cal J}_0)^2+{\cal A}_4{\cal J}_0'+{\cal A}_5{\cal J}_0+{\cal A}_6\right)=0,
 \end{gather*}
 where ${\cal A}_j={\cal A}_j({\cal H},{\cal L}_2,{\cal L}_3,{\cal K}_0)$.  Since ${\cal Q}$ factors out of the
equation, the basic identity is of order~12:
\[
{\cal A}_1({\cal J}_0')^2+{\cal A}_2{\cal J}_0'{\cal J}_0+{\cal A}_3({\cal J}_0)^2+{\cal A}_4{\cal J}_0'+{\cal A}_5{\cal J}_0+{\cal A}_6=0.
\]
The leading coef\/f\/icient is ${\cal A}_1=-4{\cal Q},$ and
\begin{gather*}
{\cal A}_2= 8{\cal L}_2 {\cal L}_3 + 2 {\cal L}_2{\cal  K}_0 + 2{\cal  K}_0 {\cal L}_3 - 4 {\cal L}_2^2  - 4 {\cal L}_3^2+ 4(-b+ c+2d){\cal L}_3+(12b-12 c+8d){\cal L}_2\\
\hphantom{{\cal A}_2=}{} - 2 d{\cal K}_0-4 cd+4 bd
- 4 d^2,
\\
{\cal A}_3= - 2{\cal L}_2{\cal L}_3 +{\cal L}_2 {\cal K}_0- {\cal L}_3^2-  {\cal L}_2^2 + {\cal K}_0{\cal  L}_3 - \frac14 {\cal K}_0^2 + 2(- b+c+d) {\cal L}_3  + 2 (7b+c+d) {\cal L}_2\\
\hphantom{{\cal A}_3=}{}
+ (b-c-d) {\cal K}_0  -b^2-c^2- d^2  + 2 b d + 2 b c - 2 cd, \\
{\cal A}_4=8a^2{\cal L}_2^2 -12 a^2 {\cal K}_0{\cal L}_3+8a^2{\cal L}_3^2-16a^2{\cal L}_2{\cal L}_3 +4 a^2 {\cal K}_0{\cal L}_2+8a^2(-  b+  c-2d){\cal L}_2\\
\hphantom{{\cal A}_4=}{}
 -4a^2 d {\cal K}_0
+8a^2(-  b+c-2d){\cal L}_3 +8a^2 d^2 -40a^2cd  +40a^2bd,\\
  {\cal A}_5=+4a^2{\cal L}_2^2+20a^2{\cal L}_3^2-a^2{\cal K}_0^2 -8a^2{\cal L}_2{\cal L}_3 -8a^2{\cal  K}_0{\cal L}_3+8a^2(-2 b+2c-d){\cal L}_2 \\
\hphantom{{\cal A}_5=}{}
   -8a^2(4b+ 4 c+3d){\cal L}_3-4 a^2( b-c){\cal K}_0 +4 a^2d^2 + 12 a^2b^2
 + 16a^2 cd  -24 a^2 bc\\
\hphantom{{\cal A}_5=}{}
 + 12 a^2 c^2 +48 a^2 bd,\\
{\cal A}_6=-4 a^4{\cal L}_2^2- 36a^4{\cal L}_3^2 + 128a^2{\cal L}_2^2{\cal L}_3{\cal H}  - 256a^2{\cal L}_2{\cal L}_3^2 {\cal H}
- 512d{\cal L}_2^2 {\cal L}_3 {\cal H}^2- 512 {\cal L}_2^2{\cal L}_3^2{\cal  H}^2\\
\hphantom{{\cal A}_6=}{}
+ 256{\cal L}_2^3{\cal L}_3{\cal H}^2 - 256a^2d{\cal L}_2{\cal L}_3{\cal  H}
 - 4a^4d^2 + 8a^4d{\cal L}_2
- 24 a^4d{\cal L}_3  + 24 a^4{\cal L}_2 {\cal L}_3
- 36 a^4 d^2 \\
\hphantom{{\cal A}_6=}{}
- 36 a^4 c^2
- 24a^4b{\cal L}_2- 40a^4c{\cal  L}_2 - 256a^2(b+c){\cal L}_2^2{\cal H}- 512( b+c){\cal L}_2^3{\cal H}^2
 + 72a^4b {\cal L}_3 \\
\hphantom{{\cal A}_6=}{}
  + 56a^4c{\cal L}_3 + 72 a^4 b c- 512(b^2 +c^2){\cal L}_2^2{\cal H}^2
+ 128a^2{\cal L}_3^3 {\cal  H} + 512 a^2(b+c){\cal L}_2{\cal  L}_3 {\cal H}  \\
\hphantom{{\cal A}_6=}{}
 + 1024(b+c) {\cal L}_2^2{\cal L}_3{\cal H}^2- 256a^2(b^2+c^2){\cal L}_2 {\cal H}+
512 a^2bc {\cal L}_2 {\cal H}+ 1024 bc{\cal L}_2^2{\cal H}^2\\
\hphantom{{\cal A}_6=}{}
- 256a^2(+c)b{\cal L}_3^2{\cal H}
 + 256{\cal L}_2 {\cal L}_3^3 {\cal H}^22 - 512(b+c){\cal L}_2{\cal L}_3^2{\cal H}^2+ 128a^2(b-c)^2{\cal L}_3{\cal H}\\
\hphantom{{\cal A}_6=}{}
  + 256(b-c)^2{\cal L}_2 {\cal L}_3 {\cal H}^2 - a^4 {\cal K}_0^2 +24 a^4b d+ 104a^4 cd+ 12 a^4( c-b){\cal K}_0
 -4a^4{\cal L}_2 {\cal  K}_0\\
\hphantom{{\cal A}_6=}{}
 + 512a^2bd{\cal L}_2{\cal  H}+ 128a^2c{\cal L}_2{\cal K}_0{\cal  H}+ 512a^2bcd{\cal H}
 + 128a^2bd{\cal K}_0{\cal H}-128a^2b{\cal L}_2{\cal K}_0{\cal H}\\
\hphantom{{\cal A}_6=}{}
 -128a^2cd{\cal K}_0{\cal H}+ 1024bcd{\cal L}_2{\cal H}^2
 + 256d(b-c){\cal L}_2{\cal K}_0{\cal H}^2+ 512a^2c{\cal L}_2{\cal H}\\
\hphantom{{\cal A}_6=}{}
 + 1024d(b+c){\cal L}_2^2{\cal H}^2
- 256a^2d(b^2+c^2+bc+cd){\cal H}
 + 256c{\cal L}^2{\cal K}_0{\cal H}^2\\
\hphantom{{\cal A}_6=}{}
 - 512d(b^2+c^2+bd+cd){\cal L}_2{\cal H}^2- 256b{\cal L}_2^2{\cal K}_0{\cal H}^2
 +4 a^4d{\cal  K}_0
- 256La^2d{\cal L}_3^2{\cal H}\\
\hphantom{{\cal A}_6=}{}
- 512d{\cal L}_2{\cal  L}_3^2{\cal  H}^2+ 128a^2d^2{\cal L}_3 {\cal H}+ 256d^2{\cal L}_2{\cal L}_3{\cal  H}^2
 - 32a^2{\cal L}_3{\cal K}_0^2{\cal H} -64{\cal L}_2{\cal L}_3{\cal K}_0^2{\cal  H}^2\\
\hphantom{{\cal A}_6=}{}
  + 12 a^4{\cal L}_3 {\cal  K}_0+ 512a^2d(b+c){\cal L}_3 {\cal H}
+ 1024d(b+c){\cal L}_2{\cal L}_3{\cal H}^2.
\end{gather*}

 By substituting (\ref{K0R11}) into expressions (\ref{ident1})--(\ref{ident4}) and  (\ref{L2R3})--(\ref{J0R1}) for $k_1=k_2=1$,  and simplifying,
 we can recast each of these relations into polynomial structure equations in the generators. Further by squaring (\ref{R3}), substituting the
structure equations for ${\cal R}_1^2$, ${\cal R}_2^2$ and ${\cal R}_1{\cal R}_2$ into the result and simplifying, we obtain the polynomial structure
equations for ${\cal R}_3^2$. Multiplying (\ref{R3}) by ${\cal R}_1$  substituting the structure equations for ${\cal R}_1^2$ and ${\cal R}_1{\cal R}_2$ into
 the result and simplifying, we obtain the polynomial structure equations for~${\cal R}_1{\cal R}_3$. Similarly, multiplying~(\ref{R3}) by~${\cal R}_2$
 we can obtain the polynomial structure equations for ${\cal R}_2{\cal R}_3$.

 This entire construction carries over easily for the primed symmetries and their basic commutators~(\ref{R'idents}). The only gap remaining to prove polynomial
 closure is consideration of the basic commutator ${\cal R}_0=\{{\cal J}_0,{\cal J}_0'\}$ and double commutators involving ${\cal J}_0$, ${\cal J}_0'$ simultaneously.
 An important observation here is that under any variable-parameter transposition ${\cal R}_0$ changes sign and ${\cal R}^2_0$
 is invariant. Using this observation, we have verif\/ied that
\begin{gather*}
{\cal R}^2_0= 65536{\cal H}^4\Big[{\cal I}_{xy}{\cal I}_{xz}{\cal I}_{yz}-\beta{\cal I}_{yz}({\cal I}_{xy}+{\cal I}_{xz})
 -\gamma{\cal I}_{xz}({\cal I}_{xy}+{\cal I}_{yz})\nonumber\\
 \hphantom{{\cal R}^2_0=}{}
 -\delta{\cal I}_{xy}({\cal I}_{xz}+{\cal I}_{yz})-\beta{\cal I}_{yz}^2-\gamma{\cal I}_{xz}^2-\delta{\cal I}_{xy}^2
 +(\beta(\beta+3\gamma+3\delta)+\gamma\delta){\cal I}_{yz}\nonumber\\
 \hphantom{{\cal R}^2_0=}{}
  +(\gamma(\gamma+3\delta+3\beta)+\delta\beta)
{\cal I}_{xz}+(\delta(\delta+3\beta+3\gamma)+\beta\gamma){\cal I}_{xy}\nonumber\\
 \hphantom{{\cal R}^2_0=}{}
 -2(\beta\gamma^2+\beta^2\gamma+\beta\delta^2+\beta^2\delta+\gamma\delta^2+\gamma^2\delta)\Big].
\end{gather*}
In terms of our standard basis this is
\begin{gather*}
 {\cal R}^2_0=4096{\cal H}^4\Big({-}{\cal K}_0^2{\cal L}_3-4\beta{\cal K}_0{\cal L}_2+4\beta\delta{\cal K}_0+4{\cal L}_3^3-4\gamma\delta{\cal K}_0+4\gamma{\cal K}_0{\cal L}_2-8\beta {\cal L}_2^2
 -8\gamma{\cal L}_2^2\nonumber\\
 \hphantom{{\cal R}^2_0=}{}
 +4{\cal L}_3{\cal L}_2^2-8{\cal L}_3^2{\cal L}_2+4\beta^2{\cal L}_3-8\beta{\cal L}_3^2+4\gamma^2{\cal L}_3-8\gamma{\cal L}_3^2-8\delta{\cal L}_3^2+4\delta^2{\cal L}_3+16\beta\gamma\delta\nonumber\\
 \hphantom{{\cal R}^2_0=}{}
  +16\gamma\delta{\cal L}_2+16\beta\delta{\cal L}_2+16\beta\gamma{\cal L}_2
-8\beta^2{\cal L}_2+16\beta{\cal L}_3{\cal L}_2-8\gamma^2{\cal L}_2 +16\gamma{\cal L}_3{\cal L}_2\nonumber\\
 \hphantom{{\cal R}^2_0=}{}
-8\delta{\cal L}_3{\cal L}_2-8\beta\gamma{\cal L}_3-8\beta62\delta+16\beta\delta{\cal L}_3-8\gamma^2\delta+16\gamma\delta{\cal L}_3-8\beta\delta^2-8\gamma\delta^2\Big).
\end{gather*}
A simple consequence is $\{{\cal L}_2,{\cal R}_0\}=0$. The expression ${\cal K}_1^2{\cal R}_0^2$ is a perfect square in the 6~gene\-ra\-tors and gives
\begin{gather*}
{\cal K}_1{\cal R}_0= -{{\cal K}_0}^{2}{\cal L}_3-4\beta{\cal K}_0{\cal L}_2+4\beta\delta{\cal K}_0
+4{\cal L}_3^3-4\gamma\delta{\cal K}_0+4{\cal K}_0\,\gamma{\cal L}_2-8{{\cal L}_2}^{2}b-8{{\cal L}_2}^{2}\gamma
\nonumber\\
\hphantom{{\cal K}_1{\cal R}_0=}{}
+4{\cal L}_3{{\cal L}_2}^{2}-8{{\cal
L}_3}^{2}{\cal L}_2+4{\cal L}_3{\beta}^{2}-8{{\cal L}_3}^{2}\beta+4{\cal L}_3
{\gamma}^{2}
-8{{\cal L}_3}^{2}\gamma-8{{\cal L}_3}^{2}\delta+4{\cal L}_3{\delta}^{2}
\nonumber\\
\hphantom{{\cal K}_1{\cal R}_0=}{}
+16 \delta\beta\gamma+16{\cal L}_2\delta\gamma+16{\cal L}_2\delta\beta+16{\cal L}_2\beta\gamma-8{\cal L}_2
{\beta}^{2}+16{\cal L}_3{\cal L}_2\beta-8{\cal L}_2{\gamma}^{2}+16{\cal L}_3{\cal L}_2\gamma\nonumber\\
\hphantom{{\cal K}_1{\cal R}_0=}{}
-8{\cal L}_3{\cal L}_2\,\delta-8{\cal L}_3\beta\gamma-8\delta{\beta}^{2}+16{\cal L}_3\delta\beta-8\delta{\gamma}^{2}
+16{\cal L}_3\delta\gamma-8{\delta}^{2}\beta-8{\delta}^{2}\gamma,
 \end{gather*}
where the sign is determined by comparing the highest order terms.
The expression ${\cal J}_1^2{\cal R}_0^2$ is not a perfect square, but we can make it so by adding an appropriate, uniquely determined, multiple of the functional relation: $({\cal J}_1{\cal R}_0)^2={\cal J}_1^2{\cal R}_0^2-64^2{\cal H}^4F$. We then obtain
\begin{gather*}
{\cal J}_1{\cal R}_0=32{\cal H}^2\Big[\big({-}2({\cal L}_2-{\cal L}_3)^2+({\cal L}_2+{\cal L}_3){\cal K}_0\big){\cal J}_0
-4({\cal L}_2-{\cal L}_3)^2{\cal J}'_0\nonumber\\
\hphantom{{\cal J}_1{\cal R}_0=}{}
+((-6\gamma+4\delta){\cal L}_2+2(-\beta+\gamma+2\delta){\cal L}_3-\delta{\cal K}_0){\cal J}_0+
((6\beta+4\delta){\cal L}_2+8\delta{\cal L}_3){\cal J}'_0\nonumber\\
\hphantom{{\cal J}_1{\cal R}_0=}{}
 +2\delta(\beta-\gamma-\delta){\cal J}_0-4\delta^2{\cal J}'_0+\alpha^2\left(4({\cal L}_2-{\cal L}_3)^2+(2{\cal L}_2-6{\cal L}_3){\cal K}_0\right)
 \nonumber\\
\hphantom{{\cal J}_1{\cal R}_0=}{}
  + \alpha^2(-4(\beta-\gamma+2\delta)({\cal L}_2+{\cal L}_3)-2\delta{\cal K}_0)+4\alpha^2\delta(5\beta-5\gamma+\delta)\Big],
\end{gather*}
where the sign is  determined by comparing highest order terms.

Note that
\begin{gather*}
\{{\cal J}_1,{\cal J}_0\}=-2{\cal J}^2_0+128{\cal H}^2\big(3{\cal L}_2^2+{\cal L}_3^2-4\delta{\cal L}_2-2\delta{\cal L}_3-4{\cal L}_2{\cal L}_3+\delta^2\big)\\
\hphantom{\{{\cal J}_1,{\cal J}_0\}=}{}
 +128\alpha^2{\cal H}({\cal L}_2-{\cal L}_3-\delta)+8\alpha^4.
 \end{gather*}

The problem of computing the double commutator $\{{\cal J}_0,{\cal R}_0\}$ is greatly simplif\/ied by noting that it changes sign under the
transposition symmetry $(x,\beta)\leftrightarrow(y,\gamma)$. We f\/ind
\begin{gather*}
\{{\cal J}_0,{\cal R}_0\} =512{\cal H}^2\Big[({\cal J}'_0{\cal I}_{yz}-{\cal J}''_0{\cal I}_{xz})+\delta({\cal J}''_0-{\cal J}'_0)
\nonumber\\
\hphantom{\{{\cal J}_0,{\cal R}_0\} =}{}
 -\gamma{\cal J}'_0+\beta{\cal J}''_0+2\alpha^2({\cal I}_{xz}-{\cal I}_{yz})-2\alpha^2(\beta-\gamma)\Big].
\end{gather*}
All other commutators and products can be obtained from the preceding results by use of the discrete transposition symmetries. Thus the symmetry algebra
closes polynomially, and the 6~ge\-ne\-ra\-tors obey a functional identity of order 12.

\section{Final comments} \label{Final Comments}

We have studied families of  Hamiltonians that generalize  the 3- and 4-parameter extended Kepler--Coulomb systems, found explicit generators for the symmetries that show these systems to be superintegrable, and worked out the explicit structure equations for the symmetry algebras determined by taking repeated Poisson brackets of the generators. We found it amazing that the structures could be computed exactly! This analysis strongly suggests that for higher order superintegrable systems in $n\ge 3$ dimensions, polynomial closure of the symmetry algebra is relatively rare and dependent on additional discrete symmetry. Rational closure seems to be common. The structure analysis shows the fundamental importance of raising and lowering symmetries for these systems, \cite{KKM10c,Marquette20103}.  These are nonpolynomial constants of the motion. However, all of the polynomial constants can be formed from them. We have no proof that the generators found by us are of minimal order in all cases. However, it is clear that all such generators must be expressible in terms of the basic rational raising and lowering symmetries.

An obvious issue is that of quantum analogs of the classical constructions. How is the problem to be quantized? What are the operator analogs of
raising and lowering symmetries and of rational closure? We will give results on this in a second paper.

In recent papers \cite{MPY2010,MPY2012} a new and very interesting approach to classical superintegrability has been developed,
based on the Galois theory for dif\/ferential equations. It remains to be understood how this approach relates to the techniques in the present paper.

\subsection*{Acknowledgement}
This work was partially supported by a grant from the Simons Foundation (\# 208754 to Willard Miller, Jr.).
\pdfbookmark[1]{References}{ref}
 \LastPageEnding


\begin{thebibliography}{99}
\footnotesize\itemsep=0pt


\bibitem{BH}
Ballesteros {\'A}., Herranz F.J., Maximal superintegrability of the generalized
  {K}epler--{C}oulomb system on {$N$}-dimensional curved spaces,
  \href{http://dx.doi.org/10.1088/1751-8113/42/24/245203}{\textit{J.~Phys.~A: Math. Theor.}} \textbf{42} (2009), 245203, 12~pages,
  \href{http://arxiv.org/abs/0903.2337}{arXiv:0903.2337}.

\bibitem{IMA}
Eastwood  M., Miller  W. (Editors),
Symmetries and overdetermined systems of partial dif\/ferential equations,
  \href{http://dx.doi.org/10.1007/978-0-387-73831-4}{\textit{The IMA Volumes in Mathematics and its Applications}}, Vol.~144,
  Springer, New York, 2008.

\bibitem{VE2008}
Evans N.W., Verrier P.E., Superintegrability of the caged anisotropic
  oscillator, \href{http://dx.doi.org/10.1063/1.2988133}{\textit{J.~Math. Phys.}} \textbf{49} (2008), 092902, 10~pages,
  \href{http://arxiv.org/abs/0808.2146}{arXiv:0808.2146}.

\bibitem{CG}
Gonera C., Note on superintegrability of TTW model, \href{http://arxiv.org/abs/1010.2915}{arXiv:1010.2915}.

\bibitem{KKM10c}
Kalnins E.G., Kress J.M., Miller W., A recurrence relation approach to higher
  order quantum superintegrability, \href{http://dx.doi.org/10.3842/SIGMA.2011.031}{\textit{SIGMA}} \textbf{7} (2011), 031,
  24~pages, \href{http://arxiv.org/abs/1011.6548}{arXiv:1011.6548}.

\bibitem{KKM10}
Kalnins E.G., Kress J.M., Miller W., Families of classical subgroup separable
  superintegrable systems, \href{http://dx.doi.org/10.1088/1751-8113/43/9/092001}{\textit{J.~Phys.~A: Math. Theor.}} \textbf{43}
  (2010), 092001, 8~pages, \href{http://arxiv.org/abs/0912.3158}{arXiv:0912.3158}.

\bibitem{KKM2007}
Kalnins E.G., Kress J.M., Miller W., Fine structure for 3{D} second-order
  superintegrable systems: three-parameter potentials, \href{http://dx.doi.org/10.1088/1751-8113/40/22/008}{\textit{J.~Phys.~A:
  Math. Theor.}} \textbf{40} (2007), 5875--5892.

\bibitem{KKM2}
Kalnins E.G., Kress J.M., Miller W., Second order superintegrable systems in
  conformally f\/lat spaces. {II}.~{T}he classical two-dimensional {S}t\"ackel
  transform, \href{http://dx.doi.org/10.1063/1.1894985}{\textit{J.~Math. Phys.}} \textbf{46} (2005), 053510, 15~pages.

\bibitem{KKM10a}
Kalnins E.G., Kress J.M., Miller W., Superintegrability and higher order
  integrals for quantum systems, \href{http://dx.doi.org/10.1088/1751-8113/43/26/265205}{\textit{J.~Phys.~A: Math. Theor.}} \textbf{43}
  (2010), 265205, 21~pages, \href{http://arxiv.org/abs/1002.2665}{arXiv:1002.2665}.

\bibitem{KKM10b}
Kalnins E.G., Kress J.M., Miller W., Tools for verifying classical and quantum
  superintegrability, \href{http://dx.doi.org/10.3842/SIGMA.2010.066}{\textit{SIGMA}} \textbf{6} (2010), 066, 23~pages,
  \href{http://arxiv.org/abs/1006.0864}{arXiv:1006.0864}.

\bibitem{KKMP2009}
Kalnins E.G., Kress J.M., Miller W., Post S., Structure theory for second order
  2{D} superintegrable systems with 1-parameter potentials, \href{http://dx.doi.org/10.3842/SIGMA.2009.008}{\textit{SIGMA}}
  \textbf{5} (2009), 008, 24~pages, \href{http://arxiv.org/abs/0901.3081}{arXiv:0901.3081}.

\bibitem{KMPog10}
Kalnins E.G., Miller W., Pogosyan G.S., Superintegrability and higher order
  constants for classical and quantum systems, \href{http://dx.doi.org/10.1134/S1063778811060159}{\textit{Phys. Atomic Nuclei}}
  \textbf{74} (2011), 914--918, \href{http://arxiv.org/abs/0912.2278}{arXiv:0912.2278}.

\bibitem{KMP10}
Kalnins E.G., Miller W., Post S., Coupling constant metamorphosis and
  {$N$}th-order symmetries in classical and quantum mechanics,
  \href{http://dx.doi.org/10.1088/1751-8113/43/3/035202}{\textit{J.~Phys.~A: Math. Theor.}} \textbf{43} (2010), 035202, 20~pages,
  \href{http://arxiv.org/abs/0908.4393}{arXiv:0908.4393}.

\bibitem{KMP}
Kalnins E.G., Miller W., Post S., Models for quadratic algebras associated with
  second order superintegrable systems in 2{D}, \href{http://dx.doi.org/10.3842/SIGMA.2008.008}{\textit{SIGMA}} \textbf{4}
  (2008), 008, 21~pages, \href{http://arxiv.org/abs/0801.2848}{arXiv:0801.2848}.


\bibitem{KMP2010}
Kalnins E.G., Miller W., Post S., Two-variable {W}ilson polynomials and the
  generic superintegrable system on the 3-sphere, \href{http://dx.doi.org/10.3842/SIGMA.2011.051}{\textit{SIGMA}} \textbf{7}
  (2011), 051, 26~pages, \href{http://arxiv.org/abs/1010.3032}{arXiv:1010.3032}.

\bibitem{MPY2010}
Maciejewski A.J., Przybylska M., Yoshida H., Necessary conditions for classical
  super-integrability of a certain family of potentials in constant curvature
  spaces, \href{http://dx.doi.org/10.1088/1751-8113/43/38/382001}{\textit{J.~Phys.~A: Math. Theor.}} \textbf{43} (2010), 382001,
  15~pages, \href{http://arxiv.org/abs/1004.3854}{arXiv:1004.3854}.

\bibitem{MPY2012}
Maciejewski A.J., Przybylska M., Yoshida H., Necessary conditions for the
  existence of additional f\/irst integrals for {H}amiltonian systems with
  homogeneous potential, \href{http://dx.doi.org/10.1088/0951-7715/25/2/255}{\textit{Nonlinearity}} \textbf{25} (2012), 255--277,
  \href{http://arxiv.org/abs/nlin.SI/0701057}{nlin.SI/0701057}.

\bibitem{Marquette20103}
Marquette I., Construction of classical superintegrable systems with higher
  order integrals of motion from ladder operators, \href{http://dx.doi.org/10.1063/1.3448925}{\textit{J.~Math. Phys.}}
  \textbf{51} (2010), 072903, 9~pages, \href{http://arxiv.org/abs/1002.3118}{arXiv:1002.3118}.

\bibitem{PW2010}
Post S., Winternitz P., An inf\/inite family of superintegrable deformations of
  the {C}oulomb potential, \href{http://dx.doi.org/10.1088/1751-8113/43/22/222001}{\textit{J.~Phys.~A: Math. Theor.}} \textbf{43}
  (2010), 222001, 11~pages, \href{http://arxiv.org/abs/1003.5230}{arXiv:1003.5230}.

\bibitem{SB2008}
Sergyeyev A., Blaszak M., Generalized {S}t\"ackel transform and reciprocal
  transformations for f\/inite-dimensional integrable systems,
  \href{http://dx.doi.org/10.1088/1751-8113/41/10/105205}{\textit{J.~Phys.~A: Math. Theor.}} \textbf{41} (2008), 105205, 20~pages,
  \href{http://arxiv.org/abs/0706.1473}{arXiv:0706.1473}.

\bibitem{DASK2011}
Tanoudis Y., Daskaloyannis C., Algebraic calculation of the energy eigenvalues
  for the nondegenerate three-dimensional {K}epler--{C}oulomb potential,
  \href{http://dx.doi.org/10.3842/SIGMA.2011.054}{\textit{SIGMA}} \textbf{7} (2011), 054, 11~pages, \href{http://arxiv.org/abs/1102.0397}{arXiv:1102.0397}.

\bibitem{SCQS}
Tempesta P., Winternitz P., Harnad J., Miller W., Pogosyan G., Rodriguez M. (Editors),
Superintegrability in classical and quantum systems, \textit{CRM Proceedings
  and Lecture Notes}, Vol.~37, American Mathematical Society, Providence, RI,
  2004.

\bibitem{TTW1}
Tremblay F., Turbiner A.V., Winternitz P., An inf\/inite family of solvable and
  integrable quantum systems on a plane, \href{http://dx.doi.org/10.1088/1751-8113/42/24/242001}{\textit{J.~Phys.~A: Math. Theor.}}
  \textbf{42} (2009), 242001, 10~pages, \href{http://arxiv.org/abs/0904.0738}{arXiv:0904.0738}.

\bibitem{TTW2}
Tremblay F., Turbiner A.V., Winternitz P., Periodic orbits for an inf\/inite
  family of classical superintegrable systems, \href{http://dx.doi.org/10.1088/1751-8113/43/1/015202}{\textit{J.~Phys.~A: Math.
  Theor.}} \textbf{43} (2010), 015202, 14~pages, \href{http://arxiv.org/abs/0910.0299}{arXiv:0910.0299}.

\bibitem{Tsiganov2008a}
Tsiganov A.V., Addition theorems and the {D}rach superintegrable systems,
  \href{http://dx.doi.org/10.1088/1751-8113/41/33/335204}{\textit{J.~Phys.~A: Math. Theor.}} \textbf{41} (2008), 335204, 16~pages,
  \href{http://arxiv.org/abs/0805.3443}{arXiv:0805.3443}.

\bibitem{Tsiganov2009}
Tsiganov A.V., Leonard {E}uler: addition theorems and superintegrable systems,
  \href{http://dx.doi.org/10.1134/S1560354709030034}{\textit{Regul. Chaotic Dyn.}} \textbf{14} (2009), 389--406,
  \href{http://arxiv.org/abs/0810.1100}{arXiv:0810.1100}.

\bibitem{Tsiganov2008}
Tsiganov A.V., On maximally superintegrable systems, \href{http://dx.doi.org/10.1134/S1560354708030040}{\textit{Regul. Chaotic
  Dyn.}} \textbf{13} (2008), 178--190, \href{http://arxiv.org/abs/0711.2225}{arXiv:0711.2225}.

\bibitem{Evans2008a}
Verrier P.E., Evans N.W., A new superintegrable {H}amiltonian, \href{http://dx.doi.org/10.1063/1.2840465}{\textit{J.~Math.
  Phys.}} \textbf{49} (2008), 022902, 8~pages, \href{http://arxiv.org/abs/0712.3677}{arXiv:0712.3677}.

\end{thebibliography}
\end{document}